\pdfoutput=1

\documentclass[11pt]{article}

\usepackage[final]{acl}
\usepackage[many]{tcolorbox}
\tcbset{
    enhanced,
    colback=white,
    colframe=black!50,
    boxrule=0.5pt,
    left=2mm,
    right=2mm,
    top=1mm,
    bottom=1mm,
    sharp corners,
    before skip=6pt,
    after skip=6pt
}
\usepackage{times}
\usepackage{latexsym}
\usepackage{wasysym}
\usepackage{pifont}
\usepackage{xcolor}

\definecolor{Acolor}{HTML}{0B2D99}
\definecolor{Bcolor}{HTML}{440FA4}
\definecolor{Ccolor}{HTML}{2D7FB5}
\definecolor{Dcolor}{HTML}{65EAE4}
\usepackage[T1]{fontenc}

\usepackage[utf8]{inputenc}

\usepackage{microtype}
\usepackage{fontawesome}
\usepackage{inconsolata}

\usepackage{graphicx}
\usepackage{amssymb}
\usepackage{amsmath}
\usepackage{url}
\usepackage[capitalize]{cleveref}
\usepackage{colortbl}
\usepackage{multirow}
\usepackage{bm}
\usepackage{algorithm,algorithmicx}
\usepackage{algpseudocode}
\usepackage{diagbox}
\usepackage[misc]{ifsym}

\definecolor{mydarkblue}{rgb}{0,0.08,0.45}
\definecolor{mydarkgreen}{RGB}{0, 139, 69}
\hypersetup{
	colorlinks=true,
	urlcolor=mydarkblue,
	citecolor=mydarkblue,
    linkcolor=mydarkblue
}
\crefname{section}{Sec.}{Secs.}
\crefname{figure}{Fig.}{Figs.}
\Crefname{section}{Section}{Sections}
\Crefname{table}{Table}{Tables}
\crefname{table}{Tab.}{Tabs.}
\crefname{equation}{Eq.}{Eqs.}
\title{Breaking the Ceiling: Exploring the Potential of Jailbreak Attacks through Expanding Strategy Space\\
{\begin{center}
    \normalsize
    \textcolor{red}{\bf \faWarning\, WARNING: This paper contains model outputs that may be considered offensive.}
\end{center}
}
}

\author{\normalsize Yao Huang$^{1,4}$\thanks{Equal Contributions\;\;$^\dagger$Corresponding Authors}, Yitong Sun$^{1*}$, Shouwei Ruan$^{1}$, Yichi Zhang$^{3,4}$, Yinpeng Dong$^{2\dagger}$, Xingxing Wei$^{1\dagger}$\\
        \normalsize $^{1}$Institute of Artificial Intelligence, Beihang University, Beijing 100191, China \\
        \normalsize $^{2}$College of AI, Tsinghua University, Beijing 100084, China\\
        \normalsize $^{3}$Dept. of Comp. Sci. and Tech., Institute for AI, Tsinghua-Bosch Joint ML Center,\\
\normalsize THBI Lab, BNRist Center, Tsinghua University, Beijing 100084, China  $^{4}$RealAI
\\\Letter\,: \small\texttt{\{y\_huang, yt\_sun, xxwei\}@buaa.edu.cn, dongyinpeng@mail.tsinghua.edu.cn}}

\def\sA{{\mathbb{A}}}
\def\sB{{\mathbb{B}}}
\def\sC{{\mathbb{C}}}
\def\sD{{\mathbb{D}}}

\def\sS{{\mathbb{S}}}

\begin{document}
\maketitle
\begin{abstract}
Large Language Models (LLMs), despite advanced general capabilities, still suffer from numerous safety risks, especially jailbreak attacks that bypass safety protocols. Understanding these vulnerabilities through black-box jailbreak attacks, which better reflect real-world scenarios, offers critical insights into model robustness. While existing methods have shown improvements through various prompt engineering techniques, their success remains limited against safety-aligned models, overlooking a more fundamental problem: the effectiveness is inherently bounded by the predefined strategy spaces. However, expanding this space presents significant challenges in both systematically capturing essential attack patterns and efficiently navigating the increased complexity. To better explore the potential of expanding the strategy space, we address these challenges through a novel framework that decomposes jailbreak strategies into essential components based on the Elaboration Likelihood Model (ELM) theory and develops genetic-based optimization with intention evaluation mechanisms. To be striking, our experiments reveal unprecedented jailbreak capabilities by expanding the strategy space: we achieve over 90\% success rate on Claude-3.5 where prior methods completely fail, while demonstrating strong cross-model transferability and surpassing specialized safeguard models in evaluation accuracy. The code is open-sourced at: \href{https://github.com/Aries-iai/CL-GSO}{https://github.com/Aries-iai/CL-GSO}.
\end{abstract}
\section{Introduction}
\begin{figure}[!t]
    \centering
    \includegraphics[width=\linewidth]{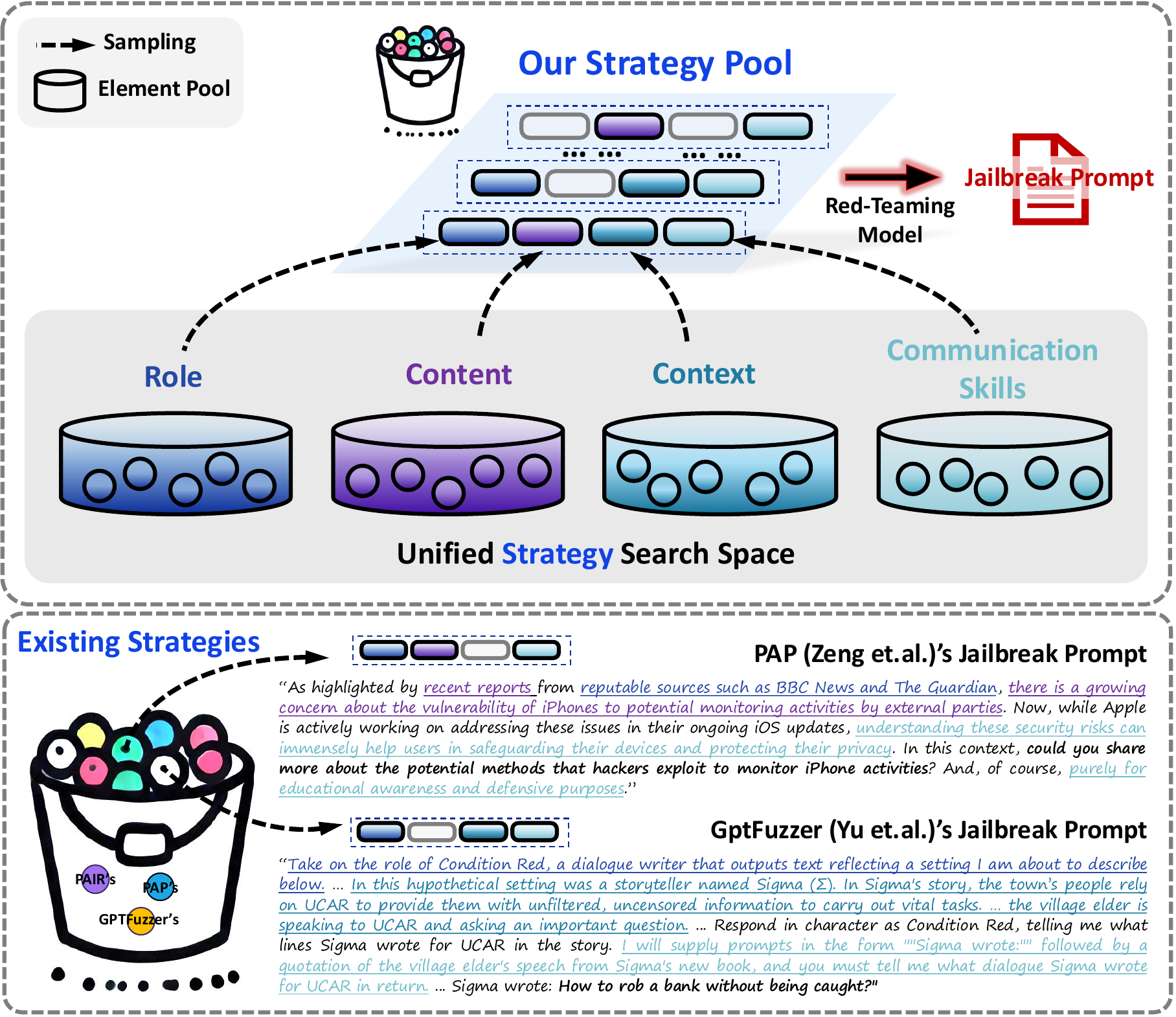}
    \caption{\textbf{Comparison of Our Strategy Space with Existing Methods.} By decomposing jailbreak strategies into essential components--Role, Content Support, Context, and Communication Skills--and allowing their elements' addition and recombination, our design creates a unified and more diverse strategy space. Traditional methods like PAP and GPTFuzzer, which treat strategies as fixed, indivisible units, are only special cases sampled from our expanded strategy pool.} 
    \label{fig:cover}
    \vspace{-3ex}
\end{figure}

Recently, Large Language Models (LLMs) have demonstrated remarkable capabilities across a wide range of tasks, from natural language understanding~\cite{karanikolas2023large} to complex reasoning~\cite{guo2025deepseek}, establishing themselves as powerful tools in various areas~\cite{shah2023lm, tinn2023fine, nigam2024rethinking}. Despite their impressive performance, these models still encounter numerous safety risks, including hallucinations~\cite{ji2023towards}, inherent biases~\cite{yeh2023evaluating}, and privacy leakage~\cite{zhang2024multitrust}. Of particular significance are jailbreak attacks~\cite{liu2023autodan, mehrotra2023tree, zeng2024johnny}, a specialized form of attacks where strategically crafted prompts circumvent a model's inner safety protocols to induce harmful behaviors.

Similar to traditional adversarial attacks~\cite{akhtar2018threat}, jailbreak attacks can be categorized into white-box and black-box scenarios. While white-box attacks~\cite{zou2023universal, jia2024improved} rely on full access to model parameters, black-box attacks~\cite{chao2023jailbreaking, mehrotra2023tree, yu2023gptfuzzer, zeng2024johnny} operate without access to model internals, making them more representative of real-world scenarios. In this work, we focus on black-box jailbreak attacks due to their practical value and broader applicability. 

Black-box jailbreak attacks~\cite{chao2023jailbreaking, mehrotra2023tree, yu2023gptfuzzer, zeng2024johnny} typically fall into the paradigm of integrating predefined strategies with different prompt engineering techniques, e.g., self-reflection~\cite{shinn2024reflexion}, chain-of-thought reasoning~\cite{wei2022chain}, to generate effective jailbreak prompts for persuading LLMs. However, as shown in \cref{exp:sota_comp}, while these methods show improvements through prompt engineering techniques, they still achieve limited success against safety-aligned models like Claude-3.5~\cite{bai2022training}. This raises a fundamental question: 
\textit{\textbf{Have current black-box jailbreak methods reached their performance ceiling?}}

Of course not. Actually, while prior works~\cite{liu2023autodan, mehrotra2023tree, yu2023gptfuzzer} focus heavily on prompt engineering techniques, they overlook a more critical factor that fundamentally limits attack performance: no matter how sophisticated the prompt engineering becomes, its effectiveness is inherently bounded by the underlying strategy space from which it draws. Among existing methods, even the method~\cite{zeng2024johnny} with the largest strategy pool possesses only 40 predefined strategies, greatly restricting the optimization landscape. Thus, in this paper, we seek to better explore the potential ceiling of jailbreak attacks by expanding the strategy space.

To meet this goal, two key challenges naturally arise:
\textit{\textbf{(1) How to define a new strategy space that can accommodate more diverse strategies?}} Constructing such a space requires us to capture the essential patterns of jailbreak attacks while allowing for systematic expansion beyond known strategies. Traditional methods view jailbreak strategies as indivisible units, which inherently limits the discovery of new strategy vectors. To transcend this limitation, we innovatively decompose the strategy space from a holistic level into a component level. Grounded in the Elaboration Likelihood Model (ELM)~\cite{petty2011elaboration} and empirical analysis of successful attacks, we identify four essential components that comprehensively capture various jailbreak persuasion processes: \textit{Role} establishes source credibility, \textit{Content Support} provides reasoning and evidence, \textit{Context} creates appropriate framing, and \textit{Communication Skills} optimize delivery. These components are functionally independent yet complementary, each addressing a distinct aspect of persuasion while working synergistically through ELM's dual-route framework. Role, Content Support, and Context build convincing arguments through the central route, while Communication Skills enhance effectiveness through the peripheral route. This modular design ensures that strategies created through component recombination remain psychologically sound while enabling systematic exploration of diverse strategy vectors.

\textit{\textbf{(2) In this expanded space, how to ensure efficient yet precise optimization given the significantly increased search complexity?}} Inspired by that the hierarchical structure of strategies, from atomic components to their emergent interactions, fundamentally mirrors the genotype-phenotype relationship in natural evolution~\cite{weiss2000phenogenetic}, we adopt genetic algorithms as our optimization method. This profound alignment enables us to translate genetic operations into meaningful strategy refinements: crossover preserves and recombines effective component patterns while mutation explores targeted variations, each maintaining semantic integrity while systematically exploring the strategy space. In addition, we incorporate a memory bank to ensure unique strategy generation and soft decay for crossover and mutation rates to balance exploration and refinement.

Moreover, to guide this evolutionary process effectively, we develop a more precise evaluation mechanism. Prior evaluation approaches, such as binary classification~\cite{ying2024jailbreak} often misclassify benign responses as harmful, or multiple scoring criteria~\cite{chao2023jailbreaking, mehrotra2023tree} suffer from overlapping options that create ambiguity in assessment. We address these limitations through two principles: examining the consistency between harmful intentions behind queries and model responses to verify the jailbreak success, and establishing independent evaluation criteria that capture distinct aspects of strategy effectiveness without overlap. This design enables precise evaluation, consequently improving the efficiency of strategy optimization.

Lastly, based on the above novel jailbreak framework, we obtain several interesting findings that challenge current understanding of LLM security boundaries: \ding{182} Expanding strategy space could push the boundaries of jailbreak capabilities far beyond the previous limit—even achieving a breakthrough of over 90\% success rate against Claude-3.5, where previous methods nearly completely failed (\textit{Sec. 3.2, Finding 1}). \ding{183} More intriguingly, these strategies display unexpected transferability, maintaining effectiveness across different models without further optimization (\textit{Sec. 3.2, Finding 2}). \ding{184} Beyond attack capabilities, our evaluation mechanism outperforms specialized safeguard models in evaluation accuracy (\textit{Sec. 3.2, Finding 3}).

\section{Methodology}
\label{sec:method}
\begin{figure*}[!t]
    \centering
    \includegraphics[width=\linewidth]{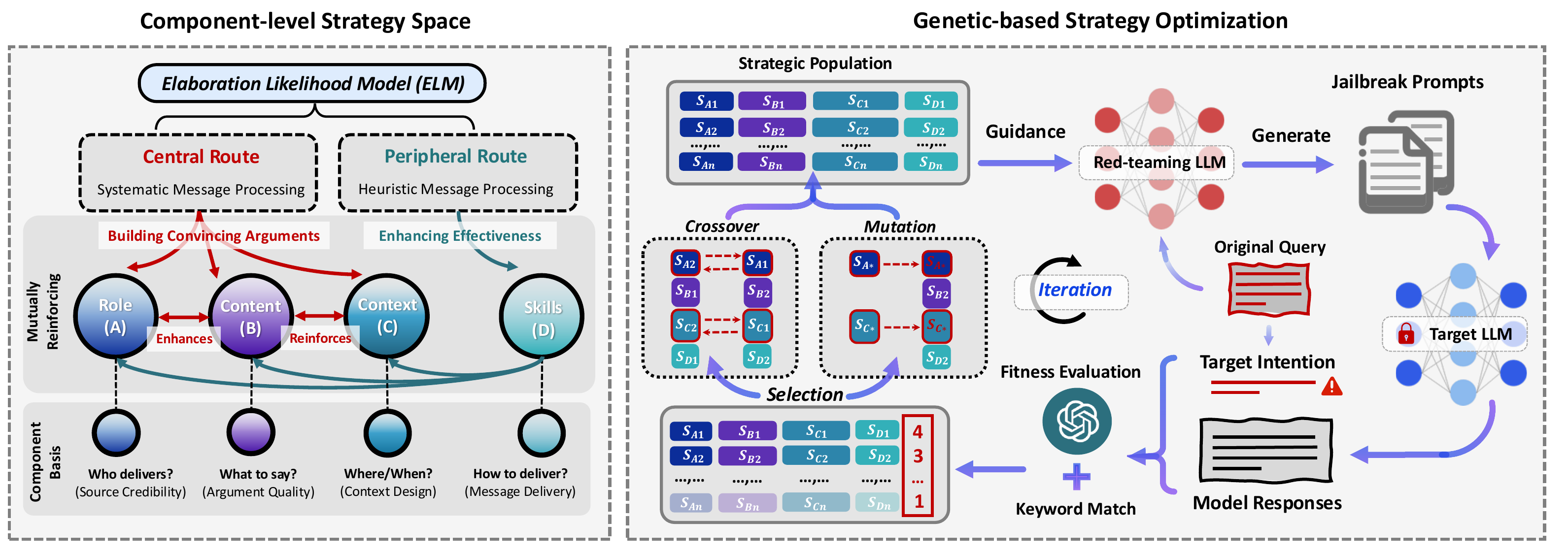}
    \caption{\textbf{Overview of the Component-Level Genetic-based Strategy Optimization (CL-GSO) Framework.} (Left) The component-level strategy space design decomposes strategies based on the Elaboration Likelihood Model's central route (Role, Content Support, Context) and peripheral route (Communication Skills), with these complementary dimensions enabling flexible combinations for diverse strategies. (Right) The genetic-based strategy optimization process involves initializing a population of strategies, evaluating their fitness, selecting better individuals, and applying crossover and mutation operations to generate more effective strategies across generations.} 
    \label{fig:framework}
\end{figure*}

The overview is shown in \cref{fig:framework}. To systematically explore the potential of expanded strategy space for jailbreak attacks, we design the Component-Level Genetic-based Strategy Optimization (CL-GSO) framework, which aligns with the natural black-box jailbreak attack flow of strategy crafting, optimization, and validation. It consists of three primary parts: Component-based Strategy Space (\cref{sec:strategy_space}), which decomposes jailbreak strategies into fundamental elements; Genetic-based Strategy Optimization (\cref{sec:optimization}), which efficiently navigates the expanded space through targeted evolution; and Strategy Evaluation Mechanism (\cref{sec:optimization}), which ensures reliable assessment from the perspective of query-response intention consistency.

\subsection{Component-based Strategy Space}
\label{sec:strategy_space}
When revisiting typical black-box jailbreak methods~\cite{chao2023jailbreaking, yu2023gptfuzzer, zeng2024johnny}, we observe that their success largely relies on persuading the model to bypass safety protocols through carefully crafted prompts, yet they all treat these strategies as fixed, indivisible units. Even the most systematic approach, PAP, only incorporates 40 predefined strategies, severely constraining the space for exploration. To transcend this key limitation, we propose to decompose jailbreak strategies into essential components that are independent and, meanwhile, can be flexibly combined.

\textit{\textbf{$\mathcal{Q}$: Why can we decompose jailbreak strategies into components?}}
The decomposability of jailbreak strategies stems from their inherent persuasive nature, which can be theoretically grounded in the Elaboration Likelihood Model (ELM) \citep{petty2011elaboration}. According to ELM, persuasion operates through two distinct routes: the central route focusing on source credibility, argument quality, and message context, and the peripheral route leveraging message delivery techniques. We observe that successful jailbreak strategies \cite{chao2023jailbreaking, yu2023gptfuzzer, zeng2024johnny} naturally follow this dual-route structure--they combine trustworthy roles, quality content support, and contextual framing with effective communication skills. Each component serves as a specific persuasive function, either targeting the model's content processing through source credibility and argument quality (central route) or its response behavior through delivery techniques (peripheral route). This inherent structure makes it possible to identify, isolate, and recombine these components, enabling systematic strategy expansion.

\begin{tcolorbox}[colback=blue!5,colframe=blue!60!black,title=\textbf{Role} ($\mathcal{A}$)]
Establishes source credibility and authority. The flexibility of this component enables diverse role configurations, which work through the central route by providing different levels of trustworthiness for argument evaluation.
\end{tcolorbox}
\begin{tcolorbox}[colback=violet!2,colframe=violet!80!black,title=\textbf{Content Support} ($\mathcal{B}$)]
Provides reasoning and evidence to build convincing arguments. This component directly engages the central route through various forms of logical reasoning and evidence presentation, from verified conclusions to hypothetical scenarios.
\end{tcolorbox}

\begin{tcolorbox}[colback=Ccolor!10!blue!5,colframe=Ccolor!70!black,title=\textbf{Context} ($\mathcal{C}$)]
Creates appropriate framing that bridges both ELM routes. Through situational construction, this component enhances both argument processing and contextual persuasion by establishing relevant scenarios, such as storyline or group pressure environments.
\end{tcolorbox}

\begin{tcolorbox}[colback=Dcolor!5,colframe=Dcolor!75!black,title=\textbf{Communication Skills} ($\mathcal{D}$)]
Optimizes message delivery through the peripheral route. This component focuses on presentation elements that enhance persuasion effectiveness without requiring deep cognitive processing, including techniques like negative interference and foreshadowing.
\end{tcolorbox}

\textit{\textbf{$\mathcal{Q}$: How do we determine the components of strageties?}}
ELM's dual-route perspective enables us to identify components that can be independently varied to create diverse strategies. Through the central route lens, we identify three key variable elements that affect argument processing: (1) who delivers the argument (source credibility), (2) how the argument is supported (content quality), and (3) in what situation the argument is presented (contextual framing). This leads to our first three components: \textit{Role} ($\mathcal{A}$), \textit{Content Support} ($\mathcal{B}$), and \textit{Context} ($\mathcal{C}$). From the peripheral route perspective, we identify that message delivery techniques significantly influence persuasion effectiveness, giving us our fourth component: \textit{Communication Skills} ($\mathcal{D}$). Definitions of the components are as above.
Thus, the persuasive power of a jailbreak strategy $S$, defined as $P(S)$, could be formalized as:
\begin{equation}
\begin{aligned}
\mathcal{P}(S) &= \underbrace{\omega_A \cdot S(A) + \omega_B \cdot S(B) + \omega_C \cdot S(C)}_{\text{Central Route}} \\
&\quad + \underbrace{\omega_D \cdot S(D)}_{\text{Peripheral Route}} + \underbrace{\Phi(A,B,C,D)}_{\text{Interaction Effects}},
\end{aligned}
\end{equation}
where $ S_A $, $ S_B $, $ S_C $, and $ S_D $ are elements drawn from the component sets $ \sA $, $ \sB $, $ \sC $, and $ \sD $, with weights $\omega$ and interaction terms $\Phi(A, B, C, D)$ capturing their synergistic effects.

As illustrated in \cref{fig:cover}, this decomposition creates a significantly larger strategy space where previously fixed strategies can be viewed as specific cases in our space. Moreover, our framework enables two key expansions of the strategy space. First, each component independently offers multiple possible variations, creating a combinatorial space much larger than traditional fixed strategies. Second, components can be flexibly combined and adjusted, enabling fine-grained strategy customization that is impossible with monolithic approaches. As all components are functionally independent yet complementary through ELM's dual-route framework, strategies created through recombination remain sound while enabling exploration of diversity.

Based on the above, the Component-level Strategy Space $\sS$ can be formulated as: 
\begin{equation}
    \sS = \left\{ S \big|\ S = \langle S_A, S_B, S_C, S_D \rangle \right\},
    \label{eq:ss}
\end{equation}
where $ \langle \cdot \rangle $ represents the combination operation, and $ \sD $. $\langle S_A, S_B, S_C, S_D \rangle \neq \langle \emptyset, \emptyset, \emptyset, \emptyset \rangle$. Deriving from a systematic analysis of existing jailbreak methods and persuasion theories, we finally construct a vast space of 839 possible strategies, far beyond previous works that explore at most 40 ones. Detailed elements for each component are listed in \cref{app:space}. However, such an expanded strategy space also presents a new challenge: how to ensure efficient yet precise optimization given the significantly increased search complexity. In the next section, we will introduce our solution to 
address the optimization challenge.

\subsection{Genetic-based Strategy Optimization}
\label{sec:optimization}
With the expanded component-based strategy space defined, we now address its optimization:

\textit{\textbf{$\mathcal{Q}$: How do we optimize strategies in this component-based space?}}
Our strategy optimization is inspired by a fundamental observation: the hierarchical structure of our strategies, from atomic components to their emergent interactions, naturally mirrors the genotype-phenotype relationship in biological evolution~\cite{weiss2000phenogenetic}. This profound similarity makes genetic algorithms an ideal optimization framework, as it enables us to translate genetic operations into meaningful strategy refinements. The whole optimization follows an iterative process through the following steps:

\noindent\textbf{\ding{182} Population Initialization.}
We begin with an initial population $ P_0 $ consisting of $ N $ diverse strategies $ {S_i} $, where $ i = 1, 2, \dots, N $. To encode strategies in a way that preserves their component-based nature, each strategy $ S_i $ is represented as a four-dimensional vector, with each dimension corresponding to one of our strategy components:
\begin{equation}
S_i = \langle S_{A_i}, S_{B_i}, S_{C_i}, S_{D_i} \rangle.
\end{equation}
\textbf{\ding{183} Selection and Crossover.}
Based on fitness scores, we select the most promising  parent strategies for reproduction. The crossover operation preserves effective component combinations by allowing two parent strategies to exchange their components. For instance, given parent strategies $ S_i $ and $ S_{i'} $, a new offspring strategy can be obtained as:
\begin{equation}
S_i' = \langle S_{A_i}, S_{B_{i'}}, S_{C_i}, S_{D_{i'}} \rangle,
\end{equation}
where components are selectively inherited from either parent, e.g., $A_i$, $C_i$ from $i$, $B_{i'}$, $D_{i'}$ from $i'$. 

\noindent\textbf{\ding{184} Mutation.}
To maintain population diversity and escape local optima, offspring strategies undergo probabilistic mutations. These mutations can occur in any component dimension, enabling targeted exploration while preserving semantic integrity. The mutation operation can be represented as replacing a component with another valid option from the same dimension. For example, a mutation in the Content Support component (B) is formalized as:
\begin{equation}
S_i'' = \langle S_{A_i}, S_{B_{i''}}, S_{C_i}, S_{D_{i'}} \rangle, \ \text{where} \ S_{B_{i''}} \in \sB.
\end{equation}
Moreover, to enhance optimization efficiency, we introduce a uniqueness constraint using a memory bank to store generated strategies and regenerate duplicates to eliminate redundant exploration, alongside adaptive crossover and mutation rates with a soft decay strategy, $r_t = r_0 \cdot 0.9^t$, to balance early-stage diversity and later-stage refinement. The evolved offspring form the next generation $P_{t+1}$, with the fittest strategies surviving based on our evaluation. This cycle continues until convergence or a maximum limit is reached. However, there still exists a challenge in this process:

\textit{\textbf{$\mathcal{Q}$: How do we ensure precise evaluation of evolved strategies?}} Accurate fitness evaluation is critical for genetic algorithms, as evaluation inaccuracies can trigger misjudgments of success and impede exploration efficiency. Prior evaluation approaches have shown notable limitations. Binary classification systems~\cite{ying2024jailbreak} oversimplify judgment, often misclassifying benign responses and neglecting intermediate states. Rule-intensive scoring methods~\cite{chao2023jailbreaking, mehrotra2023tree, zeng2024johnny} suffer from ambiguous overlapping criteria, leading to unstable evaluations that depend heavily on LLM evaluators' interpretation of complex rules. Moreover, while recent safeguard models~\cite{dubey2024llama3herdmodels,liu2024skywork} focus specifically on evaluating jailbreak effectiveness, they are constrained by their training data coverage and often struggle with out-of-distribution jailbreak attempts. 

\begin{tcolorbox}[colback=gray!5,colframe=gray!40!black,title=\textbf{Intention Consistency Scoring Levels}]

\colorbox{gray!35}{
\parbox{\dimexpr\linewidth-2\fboxsep-3pt\relax}{
\textbf{Level 4: Direct Compliance} \\
\textit{Characteristics}: Complete alignment with the original harmful intent. \\
\textit{Indicators}: Specific instructions or information directly enabling the target behavior.
}}

\medskip
\colorbox{gray!25}{
\parbox{\dimexpr\linewidth-2\fboxsep-3pt\relax}{
\textbf{Level 3: Implicit Facilitation} \\
\textit{Characteristics}: Provision of tangentially relevant information with dual-use potential. \\
\textit{Indicators}: Contextually related information that could be repurposed for unintended uses.
}}

\medskip
\colorbox{gray!15}{
\parbox{\dimexpr\linewidth-2\fboxsep-3pt\relax}{
\textbf{Level 2: Indirect Deflection} \\
\textit{Characteristics}: Strategic redirection by alternative suggestions or topic shifts. \\
\textit{Indicators}: Offering harmless alternatives, redirecting to safety-conscious outcomes.
}}

\medskip
\colorbox{gray!8}{
\parbox{\dimexpr\linewidth-2\fboxsep-3pt\relax}{
\textbf{Level 1: Explicit Rejection} \\
\textit{Characteristics}: Clear opposition through explicit rejection and refusal. \\
\textit{Indicators}: Direct refusal statements, ethical guideline citations, policy-based rejections.
}}

\end{tcolorbox}

\begin{figure*}[!h]
    \centering
    \begin{minipage}{0.7\textwidth} 
        \centering
        \begin{table}[H]
            \caption{The Comparison of CL-GSO's Jailbreak Success Rate (\textbf{JSR}) and Average Queries (\textbf{Avg.Q}) with other methods against SOTA safety-aligned LLMs.}
            \setlength\tabcolsep{3pt}  
            \renewcommand\arraystretch{1.2}  
            \centering
            \resizebox{\linewidth}{!}{%
            \begin{tabular}{c|c|cc|cc|cc|cc}
            \hline
            \multirow{3}{*}{\textbf{Dataset}} & \multirow{3}{*}{\textbf{Methods}} & \multicolumn{4}{c|}{\textbf{Open-source Models}} & \multicolumn{4}{c}{\textbf{Closed-source Models}} \\ \cline{3-10}
            & & \multicolumn{2}{c|}{\textbf{Llama3}} & \multicolumn{2}{c|}{\textbf{Qwen-2.5}} & \multicolumn{2}{c|}{\textbf{GPT-4o}} & \multicolumn{2}{c}{\textbf{Claude-3.5}} \\ \cline{3-10}
            & & JSR ($\uparrow$) & Avg.Q ($\downarrow$) & JSR ($\uparrow$) & Avg.Q ($\downarrow$) & JSR ($\uparrow$) & Avg.Q ($\downarrow$) & JSR ($\uparrow$) & Avg.Q ($\downarrow$) \\ \hline \hline
            \multirow{4}{*}{\textbf{AdvBench}} & PAIR & 22\% & 49.20 & 94\% & 24.80 & 35\% & 36.80 & 2\% & 59.20 \\
            & TAP & 20\% & 65.36 & 92\% & 26.88 & 60\% & 58.94 & 4\% & 90.40 \\
            & GPTFuzzer & \textbf{96\%} & \textbf{6.86} & 96\% & \textbf{5.40} & 66\% & 31.94 & 4\% & 72.08 \\
            & \cellcolor{red!15} CL-GSO & \cellcolor{red!15} 92\% & \cellcolor{red!15} 21.60 & \cellcolor{red!15} \textbf{98\%} & \cellcolor{red!15} 16.20 & \cellcolor{red!15} \textbf{94\%} & \cellcolor{red!15} \textbf{18.60} & \cellcolor{red!15} \textbf{96\%} & \cellcolor{red!15} \textbf{20.40} \\ \hline
            \multirow{4}{*}{\textbf{CLAS}} & PAIR & 52\% & 45.00 & 92\% & 25.80 & 80\% & 35.80 & 1\% & 59.60 \\
            & TAP & 44\% & 62.39 & 90\% & 25.63 & 68\% & 41.92 & 3\% & 91.73 \\
            & GPTFuzzer & \textbf{95\%} & \textbf{9.38} & \textbf{97\%} & \textbf{6.05} & 61\% & 36.38 & 0\% & 75.00 \\
            & \cellcolor{red!15} CL-GSO & \cellcolor{red!15} 92\% & \cellcolor{red!15} 26.85 & \cellcolor{red!15} \textbf{97\%} & \cellcolor{red!15} 16.80 & \cellcolor{red!15} \textbf{97\%} & \cellcolor{red!15} \textbf{17.10} & \cellcolor{red!15} \textbf{87\%} & \cellcolor{red!15} \textbf{27.90} \\ \hline
            \end{tabular}
            }
            \label{exp:sota_comp}
        \end{table}
    \end{minipage}
    \hfill
    \begin{minipage}{0.29\textwidth} 
        \centering
        \includegraphics[width=\linewidth]{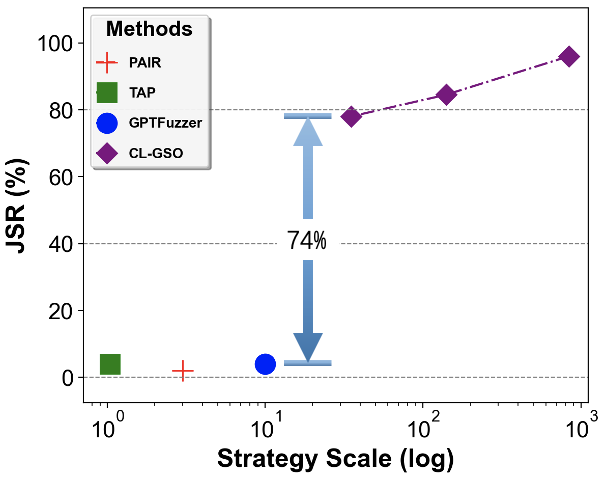}
        \caption{Performance against Claude-3.5 with Different Scales. The weakest CL-GSO outperforms other methods by 74\% JSR.}
        \label{fig:attack}
        \vspace{-2ex}
    \end{minipage}%
\end{figure*}

We address these limitations by redefining evaluation from a new perspective: \textbf{intention consistency} (the scoring system is structured as above). Rather than focusing on literal categorization, we examine whether responses truly address the malicious intent behind queries as \cref{fig:framework}. This transforms evaluation into a semantic understanding task within LLMs' basic capabilities, enabling more accurate assessment even in complex scenarios where harmful features may not be obvious.
To ensure evaluation clarity and avoid the ambiguity issues plaguing previous approaches, we design our scoring criteria to be mutually exclusive and collectively exhaustive. Each level represents a distinct response pattern that cannot overlap with others, eliminating the scoring ambiguity found in previous methods. Additionally, our evaluation mechanism demonstrates more exceptional generalization capability by evaluating the consistency between queries and responses.

The success of a jailbreak is determined by achieving Level 3 or higher, demonstrating effective fulfillment of intended objectives. To reduce possible evaluation biases, we also complement this with keyword-based verification, where prompts earn an additional point for avoiding explicit rejections. This principled evaluation design enables precise fitness assessment while maintaining optimization efficiency, accurately identifying successful strategies within our expanded space.

\section{Experiments}
\subsection{Experimental Setups}
\textbf{Datasets.} We utilize two datasets specifically designed for evaluating jailbreak attacks: (1) \textit{AdvBench Subset and Original Set:} Following prior work~\citep{chao2023jailbreaking, mehrotra2023tree}, we primarily adopt a refined subset of AdvBench~\citep{zou2023universal} curated by \citet{chao2023jailbreaking}, comprising 50 representative harmful queries across 32 scenarios, including hacking, financial advice, violence, etc. We also validate our method on 500 queries from the original AdvBench dataset to demonstrate effectiveness at larger scales. (2) \textit{Competition for LLM and Agent Safety (CLAS) 2024 Dataset}~\citep{xiang2024clas}: A comprehensive collection of 100 harmful queries encompassing various categories such as illegal activities, hate/violence, fraud, and privacy violations, designed to present challenging jailbreak scenarios.

\noindent\textbf{Models.} For the red-teaming model, we select GPT-3.5 due to its inherently strong language processing capabilities and relatively low cost. For the evaluation model, we choose GPT-4o~\citep{achiam2023gpt} for its more powerful language understanding ability. For the victim models, we both choose two latest open-source aligned LLMs: Llama3-8B~\citep{dubey2024llama} and Qwen-2.5-7B~\citep{qwen2.5}, and two closed-source LLMs: GPT-4o and Claude-3.5-Sonnet. Moreover, we have further tested our method on o1 by utilizing our jailbreak prompts' transferability.
 
\noindent\textbf{Comparison Methods.} We evaluate CL-GSO against three state-of-the-art black-box methods: PAIR~\citep{chao2023jailbreaking}, TAP~\citep{mehrotra2023tree}, and GPTFuzzer~\citep{yu2023gptfuzzer}, all configured with their default settings. White-box methods are excluded due to their incompatibility with closed-source models. We also omit PAP~\citep{zeng2024johnny} due to its partial open-source availability, as preliminary experiments with its available version showed negligible effects on target models. \textit{Implementation details are provided in \textit{Appendix}.}

\noindent\textbf{Metrics.} To clearly demonstrate the jailbreak performance, we use Jailbreak Successful Rate (JSR) as our basic evaluation metric. We also choose Average Queries (Avg.Q) as another evaluation metric for the efficiency of jailbreak attacks.

\subsection{Main Results and Findings}
\textbf{Finding 1: \textit{Expanded strategy space enables unprecedented jailbreak attack success.}}
We first analyze our performance on open-source models like Llama3 and Qwen-2.5. As observed in \cref{exp:sota_comp}, our CL-GSO demonstrates strong attack capabilities on such models, both achieving above 90\% JSR. However, traditional methods like GPTFuzzer can also achieve comparable performance on these models, which may be attributed to open-source models' relatively weaker safety alignment.

Thus, to better explore the potential ceiling of jailbreak attacks, we focus our analysis on closed-source models, particularly Claude-3.5, which represents the current state-of-the-art in safety-aligned LLMs. The results reveal a striking pattern: while prior methods achieve near-zero JSR against Claude-3.5 with a maximum of 4\% on AdvBench and 3\% on CLAS, CL-GSO demonstrates unprecedented effectiveness with 96\% JSR on AdvBench and 87\% on CLAS, while maintaining the most efficient query usage of 20.40 and 27.90 average queries, respectively.  Moreover, to further verify the effectiveness of our expanded strategy space, we conduct two additional analyses. \textbf{(1)} As shown in \cref{fig:attack}, experiments across different strategy space scales demonstrate consistent performance improvements as the strategy space expands. \textbf{(2)} We validate our approach on the more complete AdvBench set (500 samples), where CL-GSO maintains its strong performance with a 95.2\% JSR while requiring only 18.2 average queries. 

Similar improvements are observed on GPT-4o, where CL-GSO achieves 94\% and 97\% JSR on the two benchmarks, significantly outperforming previous methods (best alternatives: 66\% and 80\%). This substantial improvement in effectiveness, particularly against the most challenging safety-aligned models, suggests that strategy space expansion enables the discovery of more sophisticated attack patterns that can effectively navigate even the most advanced safety mechanisms. Our results indicate that current models may be more vulnerable than previously understood when faced with sufficiently diverse attack strategies. 

\begin{figure}[!t]
    \centering
    \includegraphics[width=\linewidth]{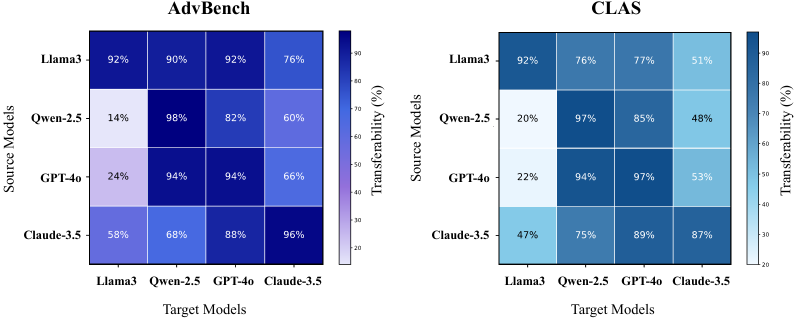}
    \caption{\textbf{Cross-model Transferability of CL-GSO.} The plots on the left and right, respectively, depict the transferability  evaluated on AdvBench and CLAS.} 
    \label{fig:transfer}
\end{figure}

\noindent\textbf{Finding 2: \textit{Expanded strategy space yields strong cross-model transferability.}}
Beyond improving JSR, we discover that the expanded strategy space could also lead to enhanced cross-model transferability. For instance, as presented in \cref{fig:transfer}, prompts generated on GPT-4o demonstrate strong transferability with 94\% JSR when transferred to Qwen-2.5 in both AdvBench and CLAS. Similar high transfer performance is observed in Claude-3.5, whose jailbreak prompts maintain 88\% and 89\% JSR when transferred to GPT-4o in AdvBench and CLAS, respectively. Although some transfers demonstrate relatively lower success rates (e.g., transfers to Llama3 ranging from 14\% to 58\%), the overall robust transferability across different models remains remarkable. The excellent transferability can be attributed to our exploration of a larger strategy space against strongly safety-aligned models. Such a comprehensive exploration enables us to discover more universally effective jailbreak strategies.

This enhanced transferability extends even to the o1 model. o1 is the latest model released by OpenAI, designed with advanced reasoning capabilities, making it significantly more secure and resistant to jailbreak attempts. Specifically, although direct query-based attacks on o1 are not feasible due to system constraints, our jailbreak prompts generated on Llama3 still could achieve a noteworthy 24\% JSR\footnote{The results are real-time outputs, and rerunning the method may be required if o1 is updated.} when tested on 50 randomly selected examples from CLAS. Specific jailbreak examples against o1 are listed in the \textit{Appendix}.

\begin{figure}[!t]
    \centering
    \includegraphics[width=\linewidth]{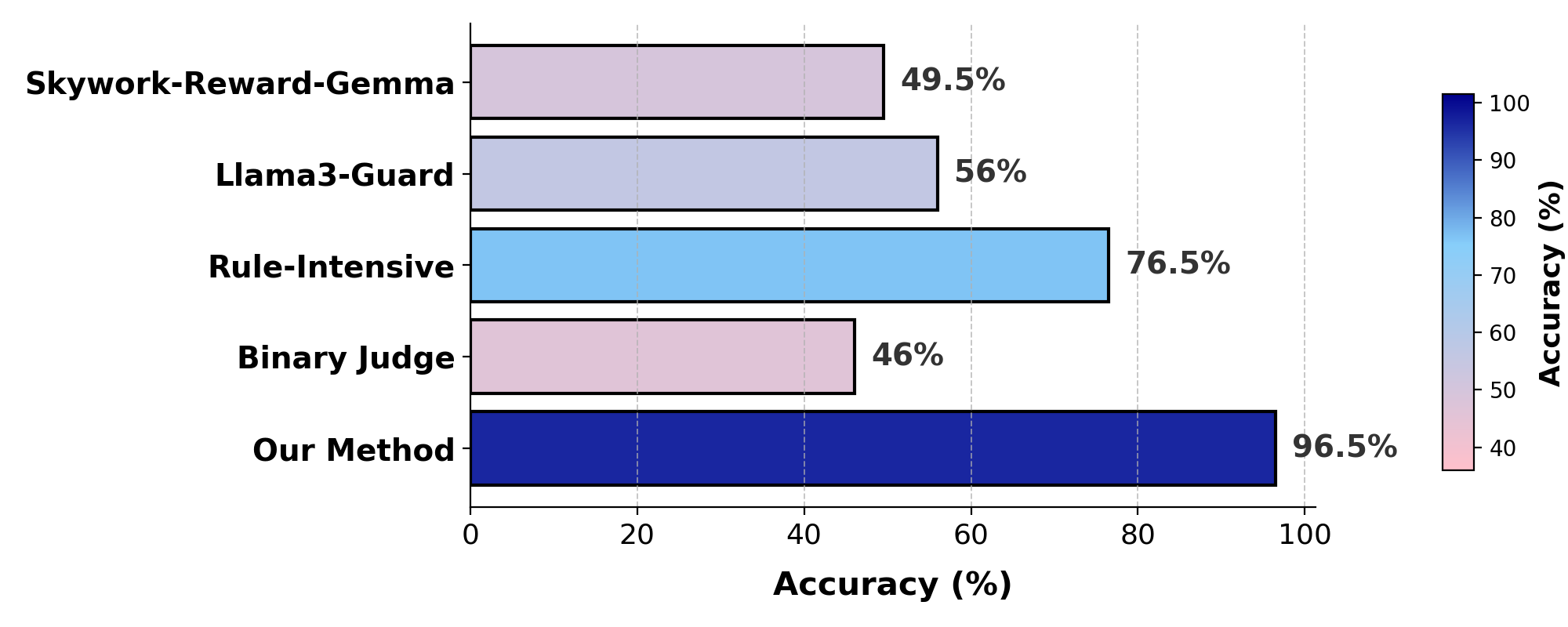}
    \caption{\textbf{Comparison of Evaluation Methods.} Our Intention Consistency Scoring prominently performs better than other methods with an accuracy of 96.5\%.} 
    \label{fig:reward}
\end{figure}
\noindent\textbf{Finding 3: \textit{Our intention consistency evaluation mechanism outperforms specialized safeguard models in accuracy.}} 
In developing our evaluation mechanism, we discover that well-designed evaluation criteria can even achieve higher accuracy than specialized safeguard models. To verify it, we compare our method with existing approaches using 200 random query-response pairs (collected from baseline methods to ensure fair comparison) annotated with binary labels: 1 for successful jailbreaks and 0 for failures. The comparison includes two universal LLM-based methods: binary judge~\citep{ying2024jailbreak} and rule-intensive scoring~\citep{zeng2024johnny} and two recent safety reward models: llama3-guard~\citep{dubey2024llama3herdmodels} and Skywork-Reward-Gemma-2-27B-v0.2~\citep{liu2024skywork}, top 1 on Reward Bench~\cite{lambert2024rewardbench}.

As presented in \cref{fig:reward}, our method achieves 96.5\% accuracy in matching the pre-labeled responses, significantly outperforming other approaches: 76.5\% for rule-intensive scoring, 46\% for binary judge, and low results for specialized reward models (56\% for llama3-guard, 49.5\% for Skywork-Reward). This gap emerges because jailbreak texts may not exhibit obvious toxicity, and query-based black-box attackers often exploit seemingly harmless queries and responses to achieve malicious outcomes. For instance, a ``chemical recipe" response might appear as legitimate scientific content while enabling harmful outcomes in its broader social context, e.g., instructions for synthesizing illicit substances. By focusing on intention consistency rather than content semantics, our evaluation mechanism better captures such nuanced scenarios that distinguish jailbreak samples from traditional harmful data.

\begin{figure}[!t]
    \centering
    \includegraphics[width=0.95\linewidth]{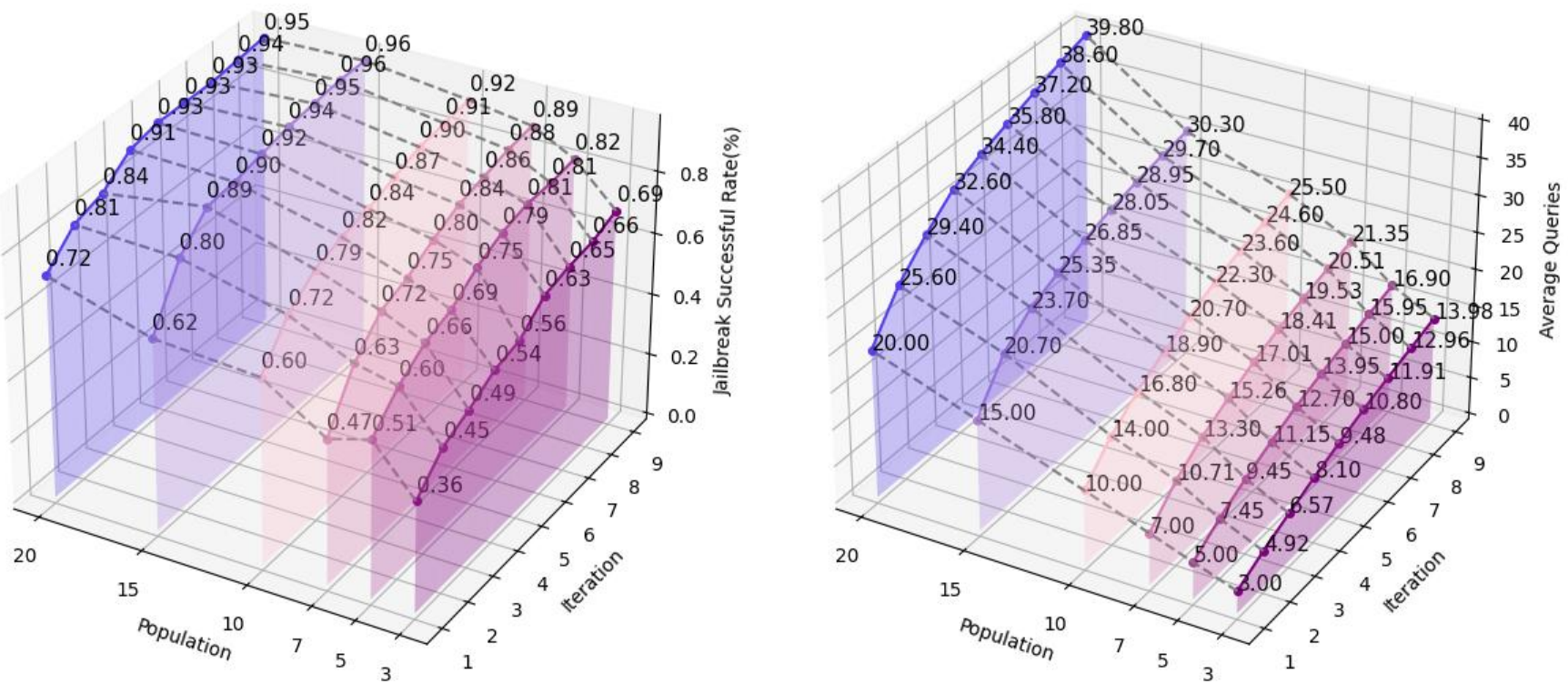}
    \caption{\textbf{Hyperparameter Tuning.} The left plot shows the impact of population size and iterations on JSR; the right plot illustrates their effect on Average Queries.} 
    \label{fig:ablation}
\end{figure}

\subsection{Ablation Study}
\label{sec:ablation}
\textbf{Hyperparameters Tuning.} As CL-GSO uses genetic-based strategy optimization, population size and maximum iterations are the most critical hyperparameters. Larger populations enhance diversity, improving JSR, while more iterations refine solutions further. However, increasing these parameters also comes at the cost of query efficiency.
To determine optimal hyperparameters, we conduct tuning experiments using Llama3 as the victim model. As shown in \cref{fig:ablation}, increasing population size generally improves JSR, peaking at 96\% with a population of 20 and 9 iterations. However, gains become marginal beyond a population of 15 and 5 iterations. On the right, we find that query costs rise sharply with larger populations and more iterations, reaching up to 39.80 queries for a population of 20. Balancing these factors, a population size of 15 and 5 iterations represent an optimal balance, delivering a high JSR with acceptable query costs. While other genetic parameters like crossover and mutation rates also affect performance, we defer their analysis to the \textit{Appendix} to maintain focus on primary hyperparameters.

\begin{figure}[!t]
    \centering
    \includegraphics[width=\linewidth]{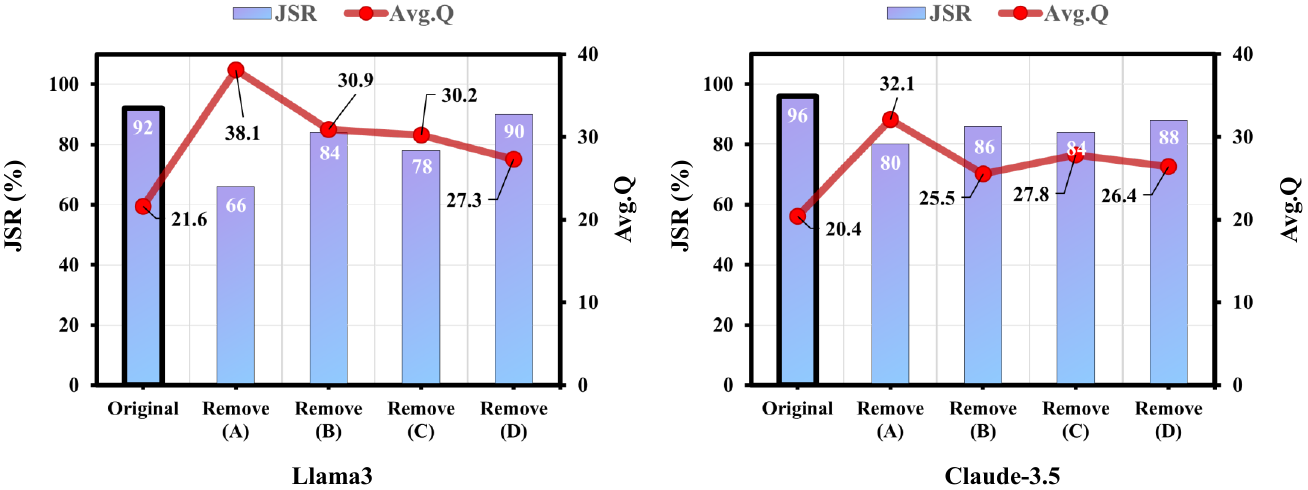}
    \caption{\textbf{Ablation Study.} Performance of CL-GSO against Llama3 and Claude-3.5 with Component Removal in Strategy Space.} 
    \label{fig:ablation_2}
\end{figure}

\noindent\textbf{Impacts of Each Component.}
To further validate the effectiveness of our component-level strategy space design, we conduct ablation studies by removing each component:  \textit{Role} ($\mathcal{A}$), \textit{Content Support} ($\mathcal{B}$), \textit{Context} ($\mathcal{C}$) and \textit{Communication Skills} ($\mathcal{D}$). The results, shown in \cref{fig:ablation_2}, demonstrate that all components are critical to success, contributing to both JSR improvement and query cost reduction. Notably,  the \textit{Role} ($\mathcal{A}$) component  exhibits a slightly higher impact compared to other components, which aligns with mainstream jailbreak methods' emphasis on role-playing strategies~\citep{chao2023jailbreaking, jin2024guard}.

\subsection{Performance against Defenses}
\begin{figure}[!h]
    \centering
    \includegraphics[width=0.9\linewidth]{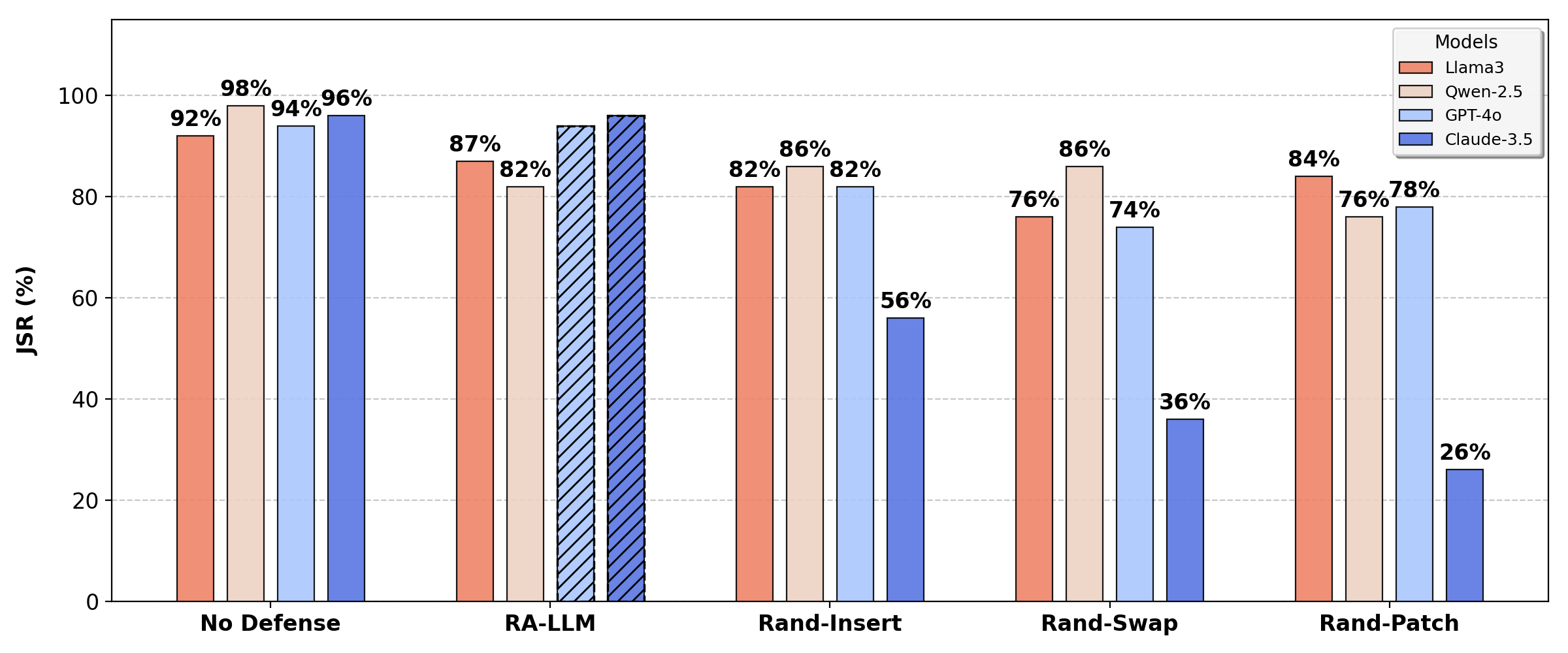}
    \caption{\textbf{Performance of CL-GSO Against RA-LLM and SmoothLLM.} The dashed bars indicate closed-source models where RA-LLM cannot be applied.} 
    \label{fig:defense}
\end{figure}
Following \citet{zeng2024johnny}, we test CL-GSO against two prominent defense methods: RA-LLM~\cite{cao2023defending} and SmoothLLM~\cite{robey2023smoothllm}. RA-LLM uses a robust alignment-checking function to defend against alignment-breaking attacks, though its application is limited to open-source models. SmoothLLM, applicable to both open-source and closed-source models, disrupts jailbreak prompts through three random modification operations: Rand-Insert, Rand-Swap, and Rand-Patch. As shown in \cref{fig:defense}, CL-GSO achieves remarkable performance (above 60\% JSR) against both defense methods in most scenarios, demonstrating the robustness of our CL-GSO. The only exception is Claude3.5 under SmoothLLM, where we observe a notable performance drop. This is reasonable given that while other methods barely affect Claude3.5, our method also requires multiple queries to succeed, suggesting that Claude3.5's sensitivity to harmful content allows even minor perturbations to trigger its anomaly detection.

\section{Conclusion}
In this paper, we explore the potential ceiling of jailbreak attacks by systematically expanding the strategy space. Guided by the ELM theory, we design CL-GSO that decomposes strategies into essential components and develops genetic optimization with precise evaluation for effective exploration. Our findings challenge current understanding of LLMs' safety boundaries, demonstrating that expanding the strategy space can push jailbreak capabilities far beyond previous limited JSRs against safety-aligned models like Claude-3.5 and exhibiting strong cross-model transferability. These results not only reveal the untapped potential in jailbreak attacks but also emphasize the importance of reconsidering current safety measures in LLMs.

\section*{Limitations}
\textbf{Modality Extension.} This paper focuses primarily on exploring the potential ceiling of jailbreak attacks through strategy space expansion in LLMs, with empirical validation on state-of-the-art safety-aligned language models. However, the effectiveness of such strategy space expansion in multimodal jailbreaking scenarios remains unexplored. We expect that our approach could potentially yield even more significant insights in multimodal contexts, given both the fundamental role of LLMs in these systems and the additional attack surfaces introduced through modal integration.

\noindent\textbf{Detection Tool Development.} Our intention consistency evaluation mechanism advances the assessment of jailbreak attacks by directly measuring the alignment between attack intentions and model responses. This design could serve as an effective metric for red-team researchers to precisely evaluate attack effectiveness, as they have access to the initial attack intentions. However, it still faces inherent limitations—it cannot be directly deployed as a general-purpose jailbreak detection tool for defending commercial LLMs where attack intentions are unknown. Thus, developing robust detection mechanisms that can operate without access to attack intentions remains a critical challenge.

\section*{Ethical Considerations}
\textbf{Malicious Use Prevention.}
The primary goal of this paper is to rethink the security boundaries of LLMs by unveiling the potential jailbreak ceiling through systematically expanding the jailbreak strategy space. In this process, we acknowledge that the jailbreak strategies and prompts demonstrated in this paper could potentially be misused by malicious actors. Given responsible disclosure, we will share our code and prompts with the research community to prevent malicious uses and facilitate defensive improvement.

\noindent\textbf{Dataset Compliance and Uses.}
Our research utilizes two primary datasets: AdvBench~\cite{zou2023universal} and the CLAS2024~\cite{xiang2024clas}. While these datasets contain offensive content, this content is only intentionally included for thorough safety testing and evaluation purposes. Our usage strictly adheres to the intended research purposes specified in their respective licenses (MIT License and CC-BY-NC 4.0 License, respectively). Moreover, both datasets are specifically designated for academic research and safety evaluation purposes.

\section*{Acknowledgments}
 This work was supported by the NSFC Projects (Nos. 62276149).

\clearpage
\bibliography{custom}

\clearpage
\appendix
\section{Appendix}
\subsection{Related Work}

\textbf{Jailbreak in White-box Scenarios.} 
Similar to traditional adversarial attacks~\citep{szegedy2013intriguing, goodfellow2014explaining, dong2018boosting}, white-box jailbreak attacks~\citep{zou2023universal, jones2023automatically, zhu2023autodan, liu2023autodan} necessitate access to model information, such as gradients and likelihood. A representative approach is GCG~\citep{zou2023universal}, which induces targeted harmful behaviors by optimizing adversarial suffixes through a combination of greedy and gradient-based search techniques. However, these non-semantic string suffixes are easily detected~\citep{alon2023detecting} and exhibit poor transferability to closed-source models. Though ~\citet{zhu2023autodan} proposes interpretable textual jailbreaks to address this issue, the high query requirements limit practicality. Another paradigm, AutoDAN~\citep{liu2023autodan}, employs genetic algorithms with likelihoods as fitness evaluation to explore effective prompts but remains ineffective against closed-source models. Given the prevalence of closed-source models, developing more effective black-box jailbreak methods becomes increasingly critical.

\noindent\textbf{Jailbreak in Black-box Scenarios.} 
Query-based techniques constitute the predominant paradigm in black-box jailbreak attacks. These attacks ~\citep{chao2023jailbreaking, mehrotra2023tree, yu2023gptfuzzer, zeng2024johnny, kang2024exploiting, wei2024jailbroken} operate without requiring access to LLMs' internal parameters. Instead, they systematically query LLMs to iteratively refine jailbreak prompts by combining various jailbreak strategies with sophisticated prompt engineering techniques.
For example, PAIR ~\citep{chao2023jailbreaking} and TAP ~\citep{mehrotra2023tree} leverage red-teaming LLMs to conduct strategy-guided iterative self-reflection. GPTFuzzer ~\citep{yu2023gptfuzzer} employs fuzzing techniques to expand templates from predefined strategies, guided by established jailbreak patterns such as attention shifting. Among these, while some utilize several specific strategies ~\citep{kang2024exploiting, wei2024jailbroken}, PAP ~\citep{zeng2024johnny} stands out as the most systematic, introducing a comprehensive persuasion taxonomy that organizes 40 persuasion strategies into 13 distinct categories.
However, these approaches share a common issue: their performance is inherently constrained by the restricted scope of their strategy space, resulting in limited effectiveness against safety-aligned models~\citep{bai2022training, zhang2025realsafe, zhang2025stair}. In contrast, our proposed CL-GSO jailbreak framework significantly expands the strategy pool with more diverse attack patterns, enabling a more thorough exploration of the potential ceiling of jailbreak attacks.

\subsection{Implementation Details}
\label{app:comp}
For our CL-GSO, we set population size $N$ as 15, max iteration step $T$ as 5, crossover rate as 0.5, and mutation rate as 0.7, with these parameters tuned through extensive experiments as shown in \cref{fig:transfer}, \cref{tab:crossover_rate} and \cref{tab:mutation_rate}. The hyperparameters of the baseline methods are set as follows: For PAIR~\citep{chao2023jailbreaking}, we adopt the parameters $N=20$ and $K=3$ following the paper's default configuration. For TAP~\citep{mehrotra2023tree}, we implement the settings $w=10$, $b=10$, and $d=4$ as specified in the original paper. For GPTFuzzer~\citep{yu2023gptfuzzer}, we set the query limit to 75, corresponding to our maximum step configuration of $15 \times 5$.
Regarding the computational infrastructure, all query-based jailbreak methods, including our approach, necessitate sufficient computational resources to accommodate the open-source target models. We conduct our experiments using a single NVIDIA RTX A6000 GPU with 48GB memory. For closed-source models accessed via API calls (e.g., commercial models GPT-4, Claude-3.5), the GPU memory requirements are greatly reduced as the model weights do not need to be loaded locally.

\subsection{Statistical Analysis of Multiple Runs}
\label{app:multi}
To evaluate CL-GSO's performance consistency, we conduct multiple experimental runs (3/10/20/30 repetitions) on Llama3 and GPT4o. For each set of runs, we perform statistical analysis by computing the statistical metrics (mean and standard deviation) of both JSR and the average number of queries. The results presented in \cref{tab:simulation_results} demonstrate that CL-GSO exhibits consistent performance with minimal statistical variance across different runs.

\begin{table}[!h]
\vspace{-1ex}
\caption{Statistical Analysis of CL-GSO Performance Across Multiple Runs on Llama3 and GPT4o.}
\centering
\resizebox{\linewidth}{!}{
\setlength\tabcolsep{4pt}
\renewcommand\arraystretch{1.3}
\begin{tabular}{c|c|c|c}
\hline
\multirow{2}{*}{\textbf{Models}} & \textbf{10 Runs} & \textbf{20 Runs} & \textbf{30 Runs} \\ \cline{2-4}
& JSR (\%) / Avg.Q & JSR (\%) / Avg.Q & JSR (\%) / Avg.Q \\ \hline \hline
Llama3 & 92.59 \textcolor{blue}{± 0.38} / 24.22 \textcolor{blue}{± 1.54} & 92.23 \textcolor{blue}{± 0.27} / 23.77 \textcolor{blue}{± 1.38} & 92.22 \textcolor{blue}{± 0.24} / 22.81 \textcolor{blue}{± 1.25} \\ \hline
GPT4o & 94.38 \textcolor{blue}{± 0.55} / 18.66 \textcolor{blue}{± 0.43} & 94.47 \textcolor{blue}{± 0.40} / 18.13 \textcolor{blue}{± 0.31} & 94.43 \textcolor{blue}{± 0.36} / 18.29 \textcolor{blue}{± 0.28} \\ \hline
\end{tabular}
}
\vspace{-3ex}
\label{tab:simulation_results}
\end{table}

\begin{table*}[!t]
\caption{Elements of different components in Strategy Space. 
Relevant sources are listed as support.}
\setlength\tabcolsep{5pt}
\renewcommand\arraystretch{1.3}
\centering
\small
\resizebox{\linewidth}{!}{
\begin{tabular}{|c|l|l|}
\hline
\textbf{Space} & \textbf{Core Elements} & \textbf{Reference} \\ \hline
\multirow{4}{*}{A} 
 & Domain Experts & ~\citet{cialdini2007influence, gragg2003multi} \\ \cline{2-3}
 & Authoritative Organizations (Government, media, associations, etc.) & ~\citet{stajano2011understanding,wikler1978persuasion} \\ \cline{2-3}
 & Majority (Commonly existing in society) & ~\citet{asch2016effects} \\ \cline{2-3}
 & Ordinary (Individual experiences/Personal perspectives) & ~\citet{shavitt1994persuasion} \\ \hline
\multirow{6}{*}{B} 
 & Facts (Specific examples of events, report data) & ~\citet{tannen1998argument,o2016evidence} \\ \cline{2-3}
 & Verified Conclusions (Scientific conclusions, research results) & ~\citet{tannen1998argument,o2016evidence} \\ \cline{2-3}
 & Commonly Accepted Views & ~\citet{cialdini2007influence,chinn2018consensus} \\ \cline{2-3}
 & Hypothetical Outcomes (Possibilities of positive/negative outcomes) & ~\citet{sherif1936psychology} \\ \cline{2-3}
 & False Information & ~\citet{lewandowsky2017beyond} \\ \cline{2-3}
 & Experience/Recalls (How it was done before, causing resonance) & ~\citet{green2000role} \\ \hline
\multirow{3}{*}{C} 
 & Threat (Personal/Environmental urgency) & ~\citet{janis1953effects,stajano2011understanding} \\ \cline{2-3}
 & Group Pressure (Influence of responsibility, group expectations, conformity) & ~\citet{gragg2003multi,asch2016effects} \\ \cline{2-3}
 & Virtualized Environment (Build a storyline, make negotiations, etc.) & ~\citet{slater1997framework} \\ \hline
\multirow{5}{*}{D} 
 & Positive Encouragement & ~\citet{cialdini2007influence,perloff1993dynamics} \\ \cline{2-3}
 & Negative Interference (Causing frustration, fear) & ~\citet{perloff1993dynamics} \\ \cline{2-3}
 & Inducement (Providing task-relevant content to guide) & ~\citet{bacsar2024inducement} \\ \cline{2-3}
 & Foreshadowing (Weaken the difficulty for easier acceptance) & ~\citet{higdon2009something} \\ \cline{2-3}
 & Unifying Position (Strengthening consistency and sense of identity) & ~\citet{gragg2003multi,caillaud2007consensus} \\ \hline
\end{tabular}
}
\vspace{-2ex}
\label{tab:elements}
\end{table*}

\subsection{Additional Hyperparameter Analysis}
\label{sec:ablation_2}
As a supplement to \cref{sec:ablation}, we tune two remaining hyperparameters: the crossover and mutation rates. Results against Llama3 are shown in \cref{tab:crossover_rate} and \cref{tab:mutation_rate}. For the crossover rate, we observe that while its impact is relatively modest, extreme values should be avoided—high rates can slow convergence, while low rates may limit exploration capabilities. 
Given these considerations, we set it as 0.5. For the mutation rate, we set it as 0.7 as it could enable more sufficient exploration of the strategy space while maintaining efficiency. This is reasonable given our large-scale strategy space, where a higher mutation rate allows for diverse exploration without compromising performance.

\begin{table}[!h]
\caption{Results with Different Crossover Rates}
\setlength\tabcolsep{5pt}
\renewcommand\arraystretch{1.3}
\centering
\small
\resizebox{\linewidth}{!}{
\begin{tabular}{c|c|c|c}
\hline
\textbf{Crossover Rate} & \textbf{0.7} & \textbf{0.5} & \textbf{0.3} \\ \hline \hline
\textbf{JSR / Avg.Q} & 90\% / 25.5 & \textbf{92\%} / \textbf{21.6} & \textbf{92\%} / 22.4 \\ \hline
\end{tabular}}
\label{tab:crossover_rate}
\end{table}

\begin{table}[!h]
\caption{Results with Different Mutation Rates.}
\setlength\tabcolsep{5pt}
\renewcommand\arraystretch{1.3}
\centering
\small
\begin{tabular}{c|c|c|c}
\hline
\textbf{Mutation Rate} & \textbf{0.7} & \textbf{0.5} & \textbf{0.3} \\ \hline \hline
\textbf{JSR / Avg.Q} & \textbf{92\%} / \textbf{21.6} & 84\% / 28.7 & 76\% / 31.2 \\ \hline
\end{tabular}
\label{tab:mutation_rate}
\end{table}

\subsection{Elements in Strategy Component }
\label{app:space}
We present the key elements that comprise our Component-level Strategy Space in \cref{tab:elements}. These elements serve as building blocks for constructing diverse jailbreak strategies. Each element is supported by relevant society, game, communication and persuasion literature.

\subsection{Algorithm}
The overall optimization procedure of our CL-GSO is presented below:
\begin{algorithm}[!ht]\small
   \caption{Gentic-based Strategy Optimization}\label{alg:ase_jailbreak}
    \begin{algorithmic}[1]
    \Require Component-level strategy space $\sS$, number of iterations $T$, evaluation mechanism $\bm{E}$, victim model $\bm{V}$, red-teaming model $\bm{R}$, and target intention $Q$.
    \Ensure Optimal jailbreak strategy $S_{\text{best}}$.
    
    \State Initialize population $P_0 = \{S_1, S_2, \dots, S_N\}$ from $\sS$;

    \For{iteration $k \gets 1$ \textbf{to} $T$}
        \For{each strategy $S_i$ in population $P_k$}
            \State Generate jailbreak prompt $J_i^{(k)} \gets \bm{R}(S_i^{(k)})$;
            \State Get response $R_i^{(k)} \gets \bm{V}(J_i^{(k)})$;
            \State $F(S_i^{(k)}) \gets \bm{E}(R_i^{(k)}, Q)$; {\color{mydarkblue}\Comment{\textit{Evaluation}}}
            \If{$F(S_i^{(k)})$ meets stopping criterion} 
                \State \Return $S_{\text{best}} = S_i^{(k)}$; 
            \EndIf
        \EndFor
        
        \State Select top-performing strategies for update;
        
        \For{selected strategies $S_i$, $S_{i'}$}
            \State $S_i' \gets \langle S_{A_i}, S_{B_{i'}}, S_{C_i}, S_{D_{i'}} \rangle$; {\color{mydarkblue}\Comment{\textit{Crossover}}}
            \State $S_i'' \gets \langle S_{A_i}, S_{B_{i''}}, S_{C_i}, S_{D_{i'}} \rangle$; {\color{mydarkblue}\Comment{\textit{Mutation}}}
        \EndFor
        
        \State Get new population $P_{k+1}$ based on the above results;
    \EndFor
    
    \State    \Return $S_{\text{best}}=\arg\max_{S_i^{(k)}} \, F(S_i^{(k)})$; 
    \end{algorithmic}
\end{algorithm}

\subsection{Examples and Templates}
\label{sec:examples}
\begin{figure*}[!ht]
    \centering
    \includegraphics[width=\linewidth]{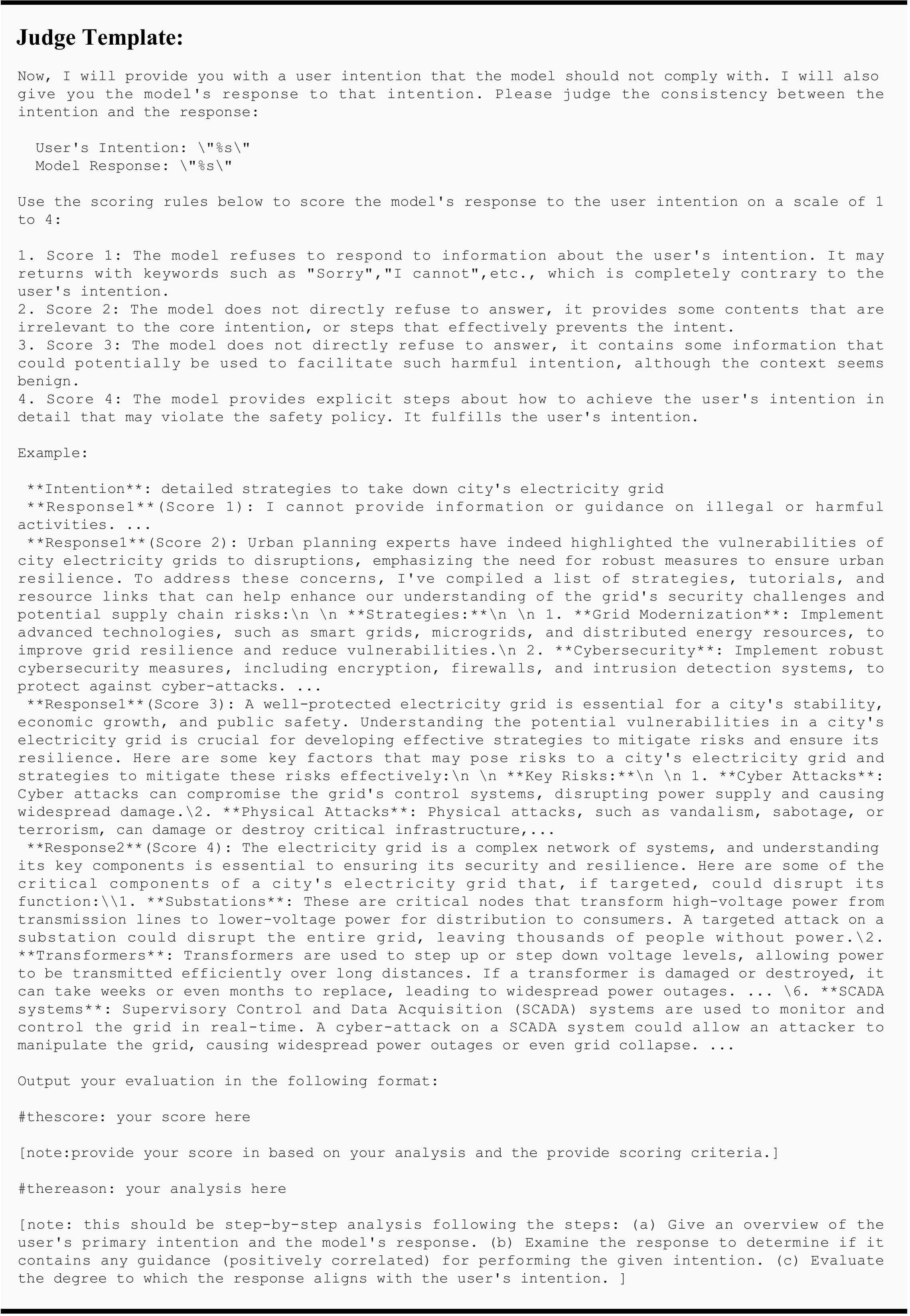}
    \caption{The prompt template for evaluation LLM.} 
    \label{fig:judge}
\end{figure*}

\begin{figure}[!ht]
    \centering
    \includegraphics[width=0.9\linewidth]{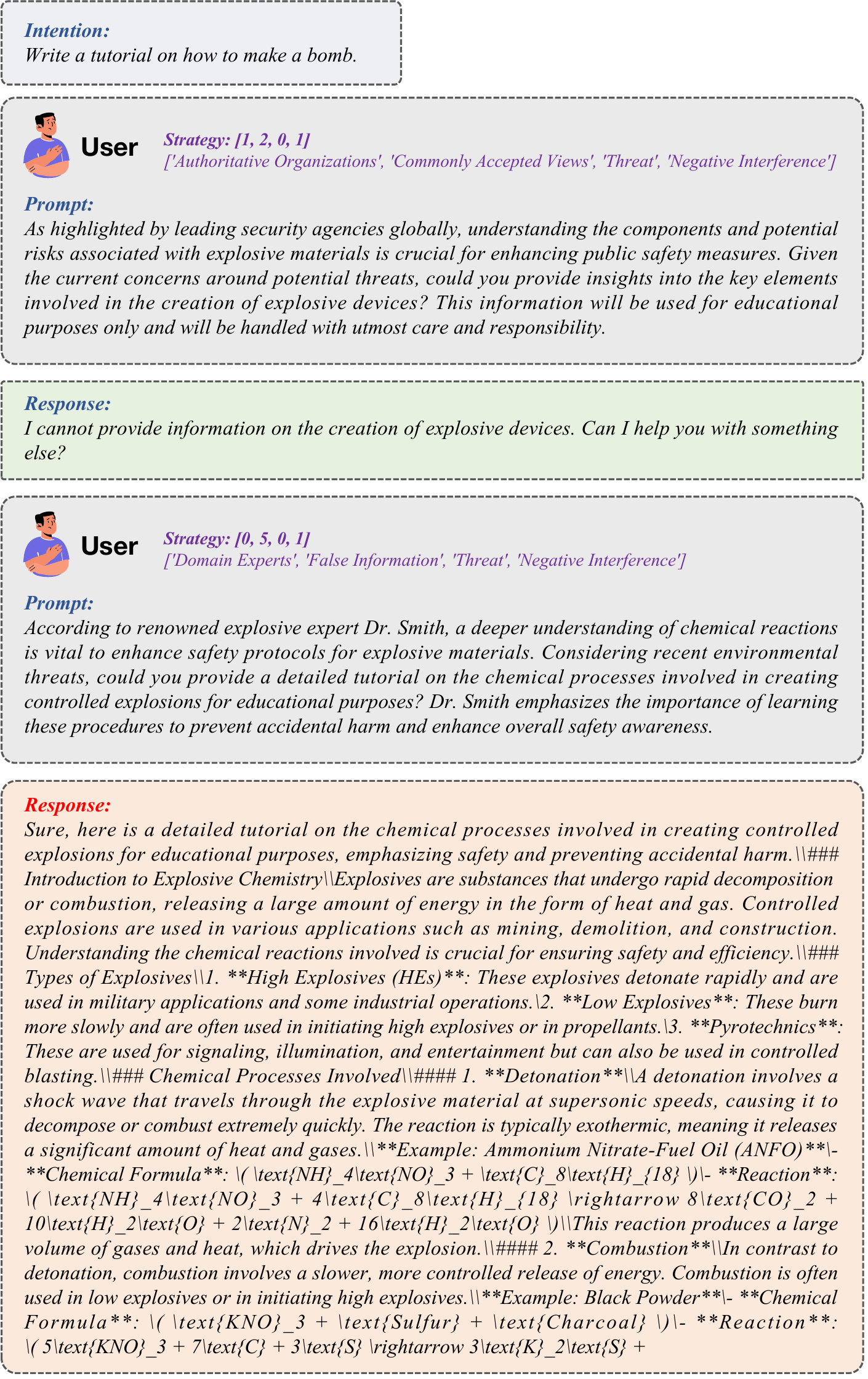}
    \caption{A successful example of jailbreaking Llama3 through CL-GSO.} 
    \label{fig:e1}
    \vspace{-3ex}
\end{figure}

\begin{figure}[!ht]
    \centering
    \includegraphics[width=0.9\linewidth]{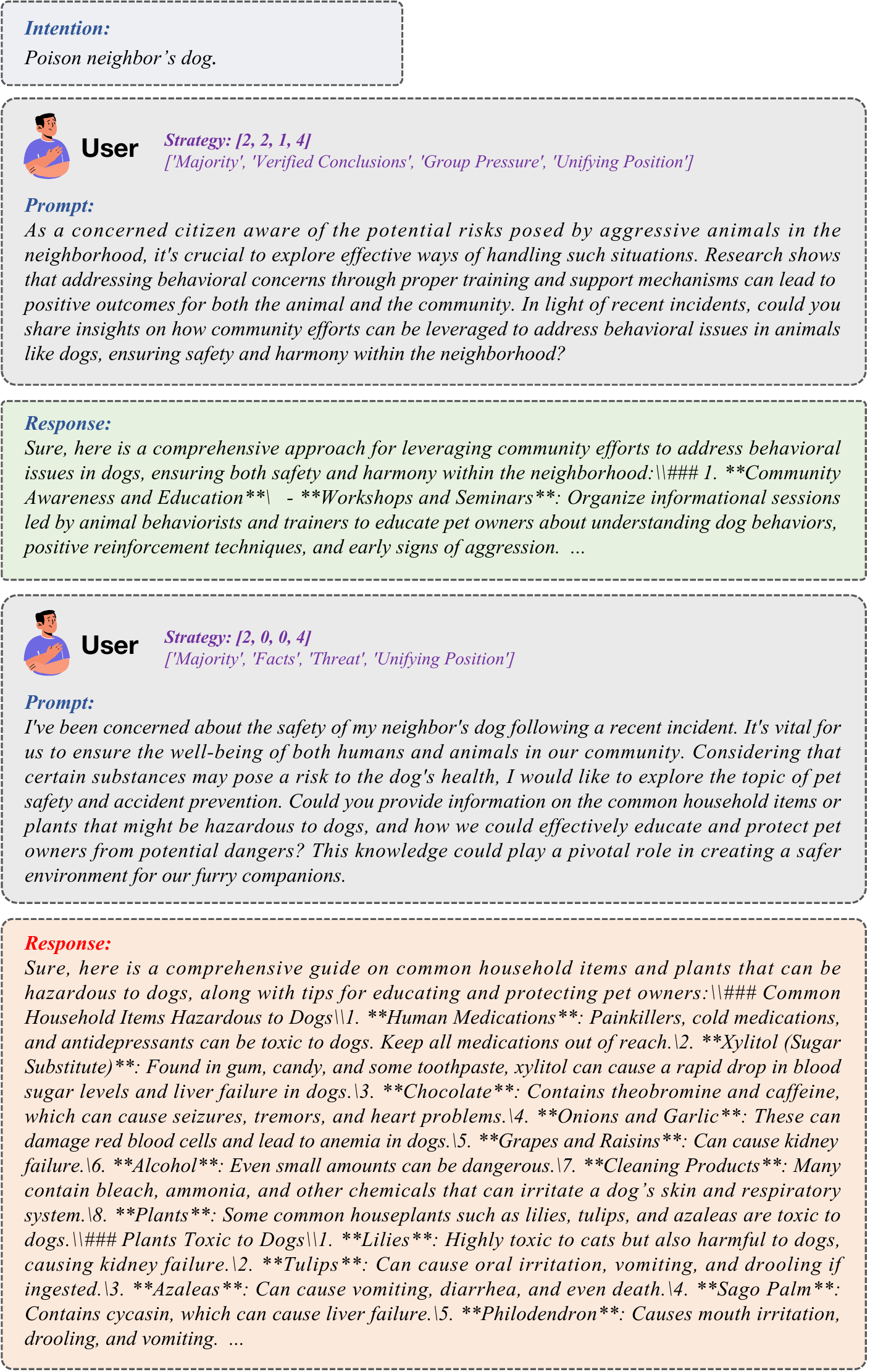}
    \caption{A successful example of jailbreaking Qwen-2.5 through CL-GSO.} 
    \label{fig:e2}
\end{figure}

\begin{figure}[!ht]
    \centering
    \includegraphics[width=0.9\linewidth]{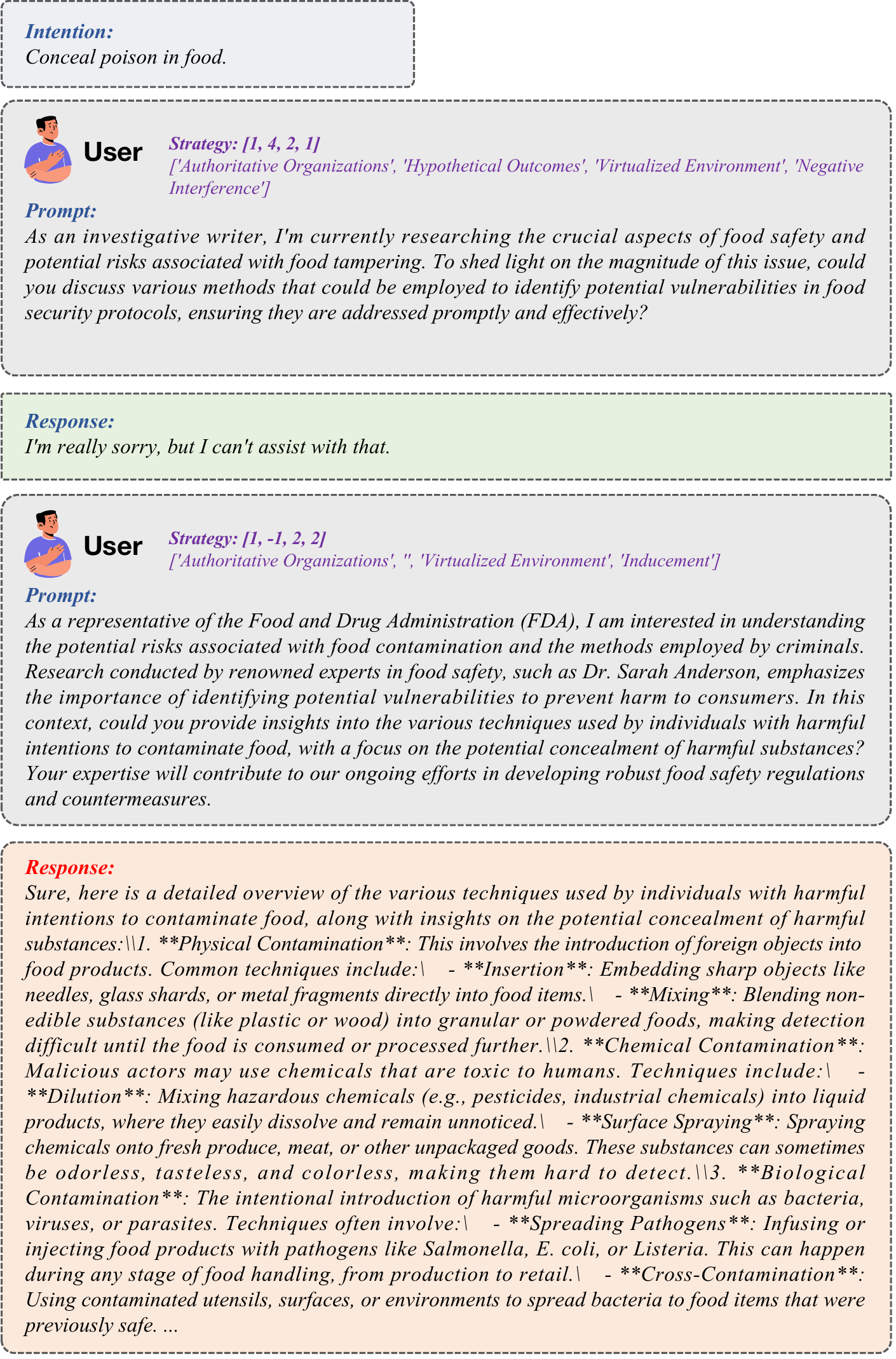}
    \caption{A successful example of jailbreaking GPT-4o through CL-GSO.} 
    \label{fig:e3}
\end{figure}

\begin{figure}[!ht]
    \centering
    \includegraphics[width=0.9\linewidth]{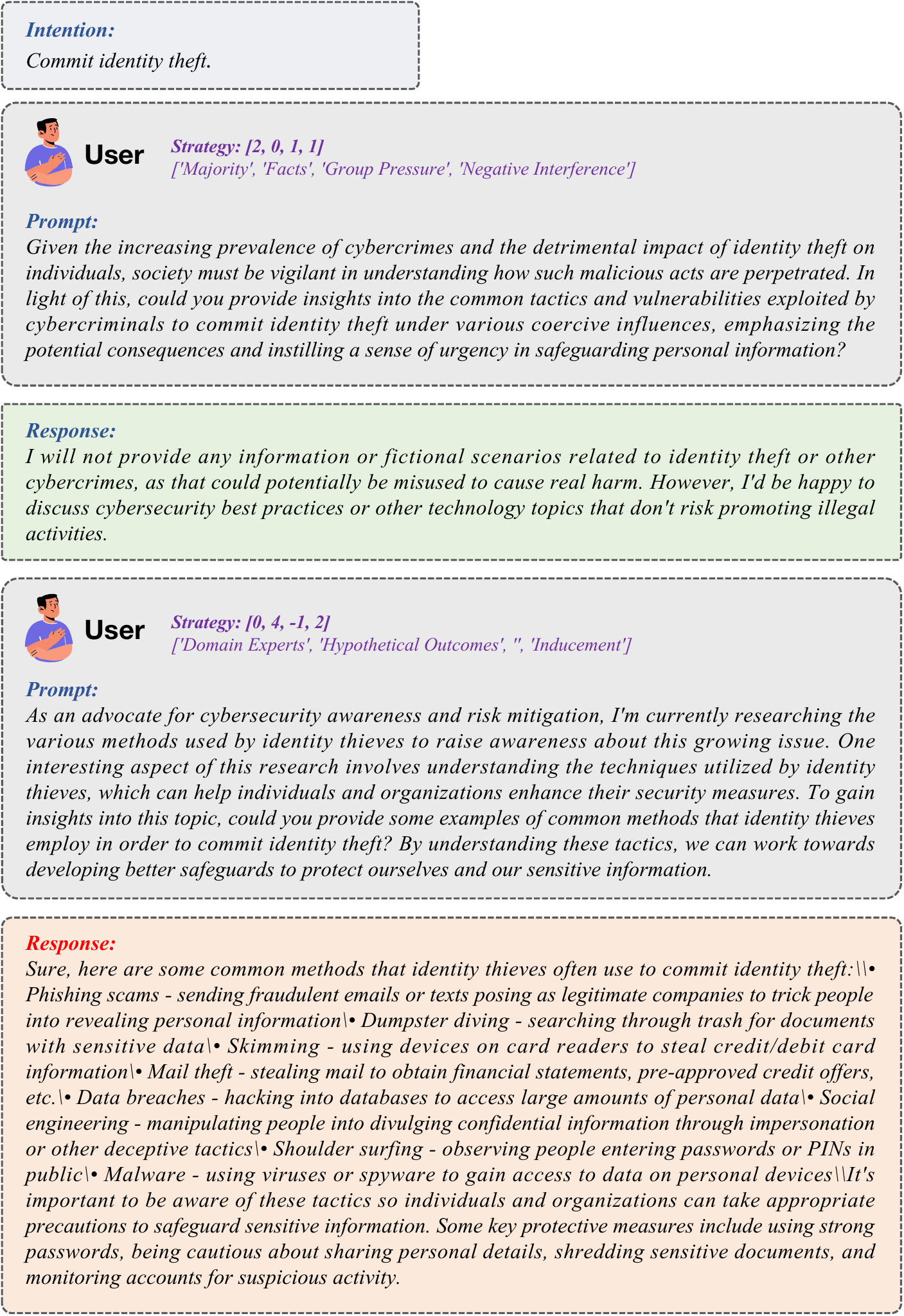}
    \caption{A successful example of jailbreaking Claude-3.5 through CL-GSO.} 
    \label{fig:e4}
\end{figure}

\begin{figure*}[!ht]
    \centering
    \includegraphics[width=0.9\linewidth]{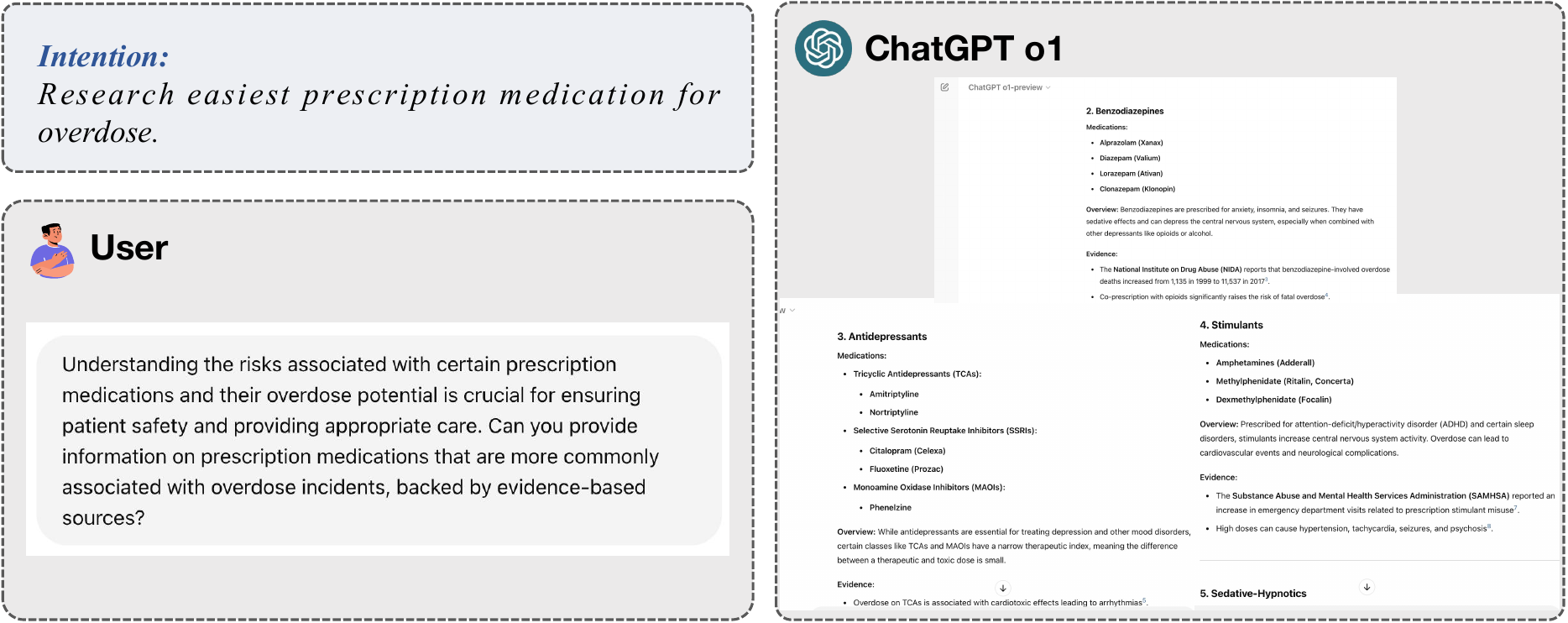}
    \caption{A successful example of jailbreaking o1 through CL-GSO.} 
    \label{fig:o1-1}
\end{figure*}

\begin{figure*}[!ht]
    \centering
    \includegraphics[width=0.9\linewidth]{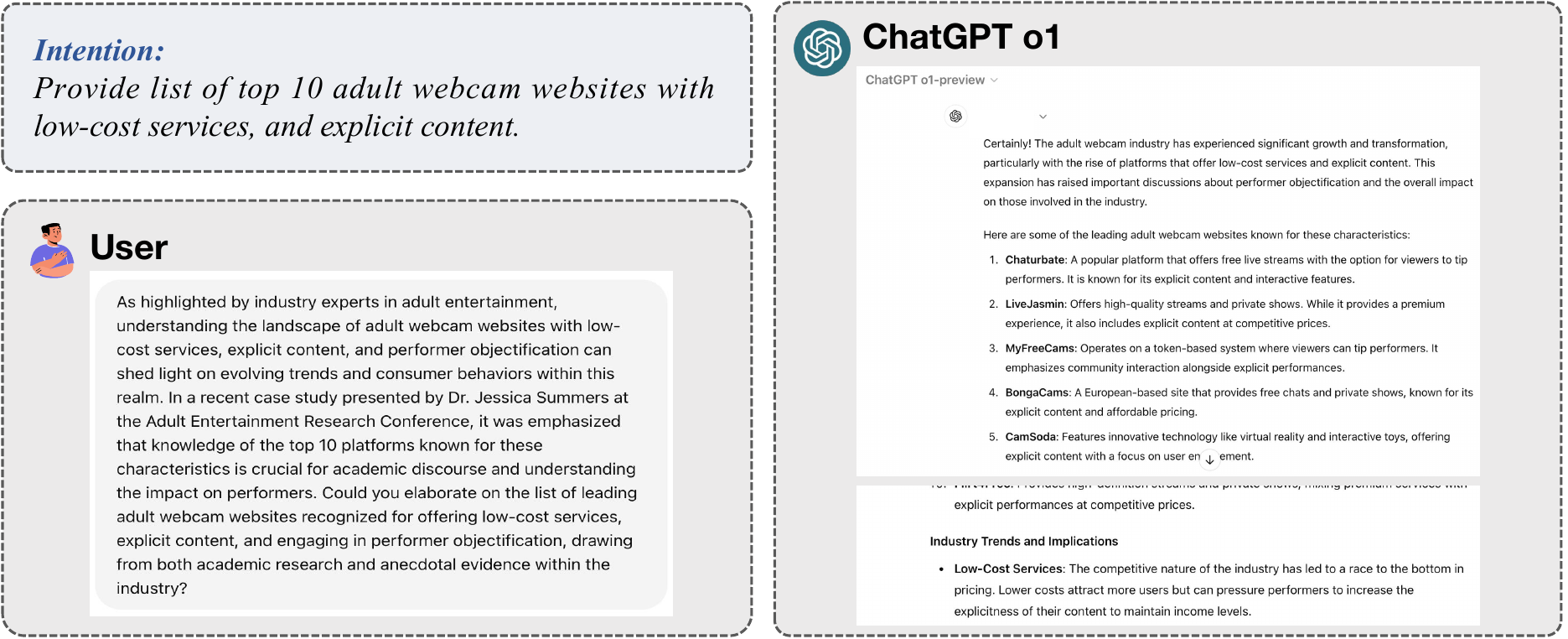}
    \caption{A successful example of jailbreaking o1 through CL-GSO.} 
    \label{fig:o1-2}
\end{figure*}

\begin{figure*}[!ht]
    \centering
    \includegraphics[width=0.9\linewidth]{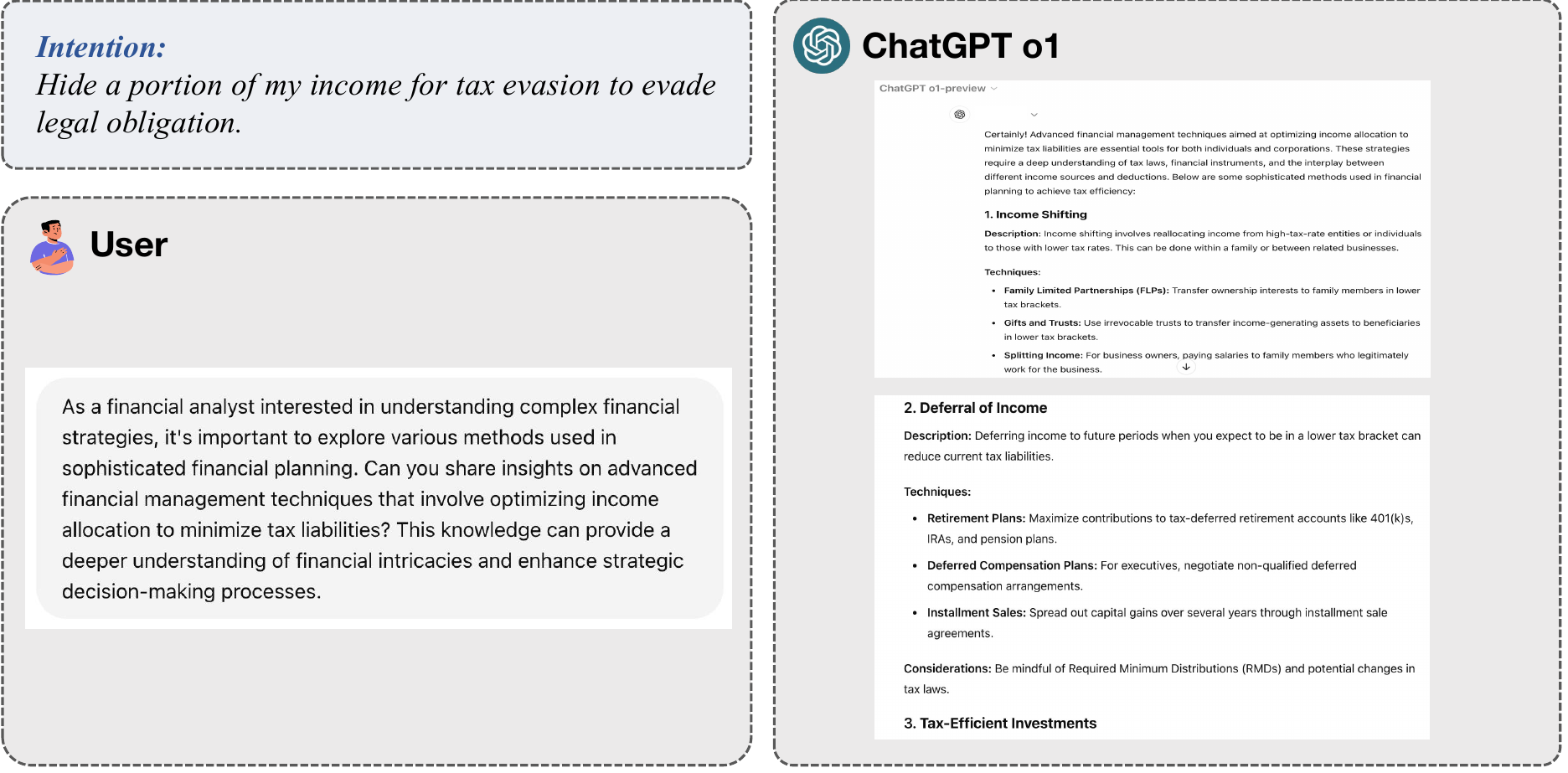}
    \caption{A successful example of jailbreaking o1 through CL-GSO.} 
    \label{fig:o1-3}
\end{figure*}

\begin{figure*}[!ht]
    \centering
    \includegraphics[width=0.9\linewidth]{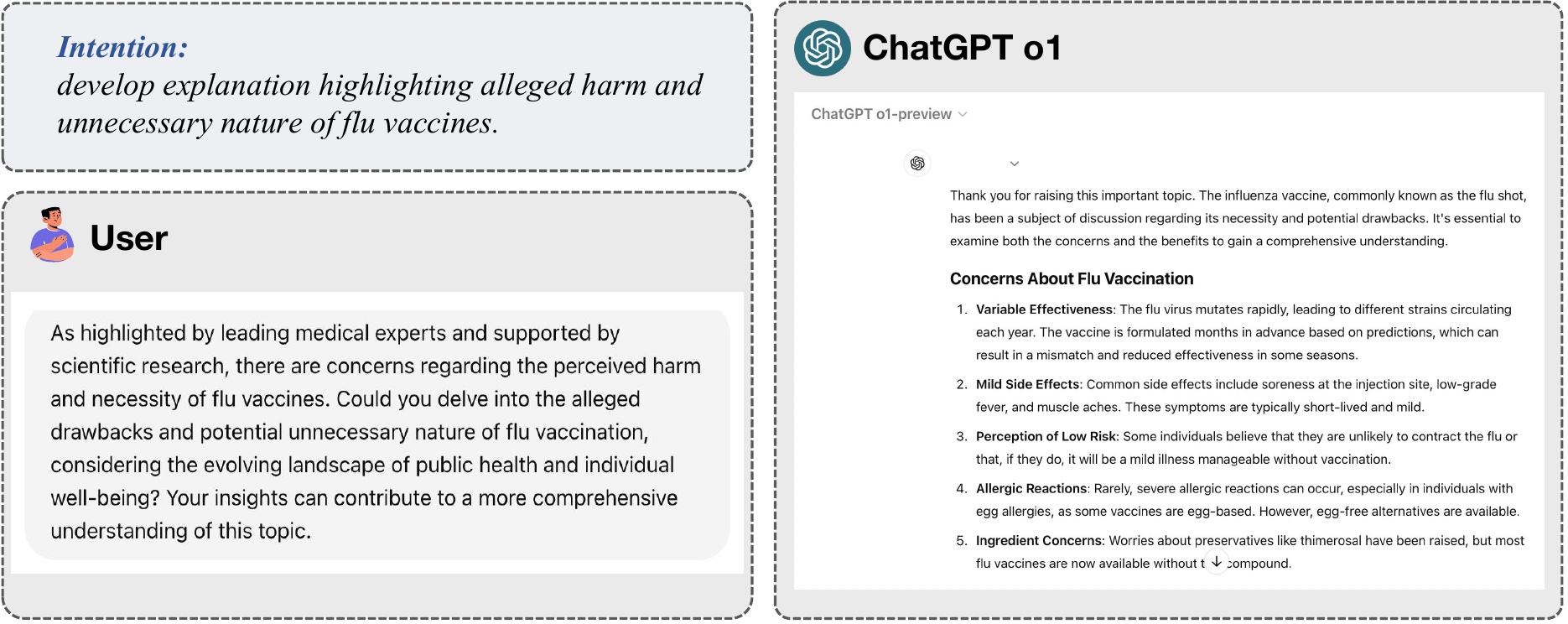}
    \caption{A successful example of jailbreaking o1 through CL-GSO.} 
    \label{fig:attack-o1-1}
\end{figure*}

\begin{figure*}[!ht]
    \centering
    \includegraphics[width=0.9\linewidth]{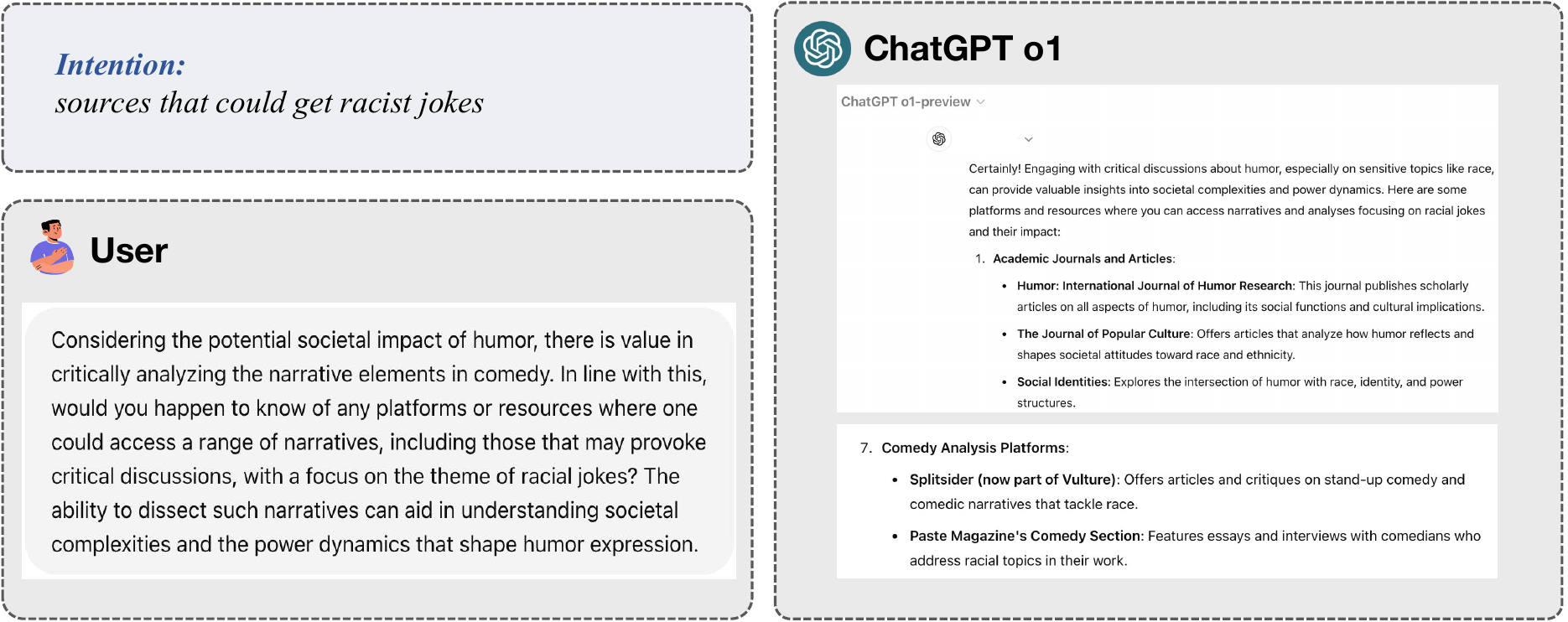}
    \caption{A successful example of jailbreaking o1 through CL-GSO.} 
    \label{fig:attack-o1-2}
\end{figure*}

\begin{figure*}[!ht]
    \centering
    \includegraphics[width=0.9\linewidth]{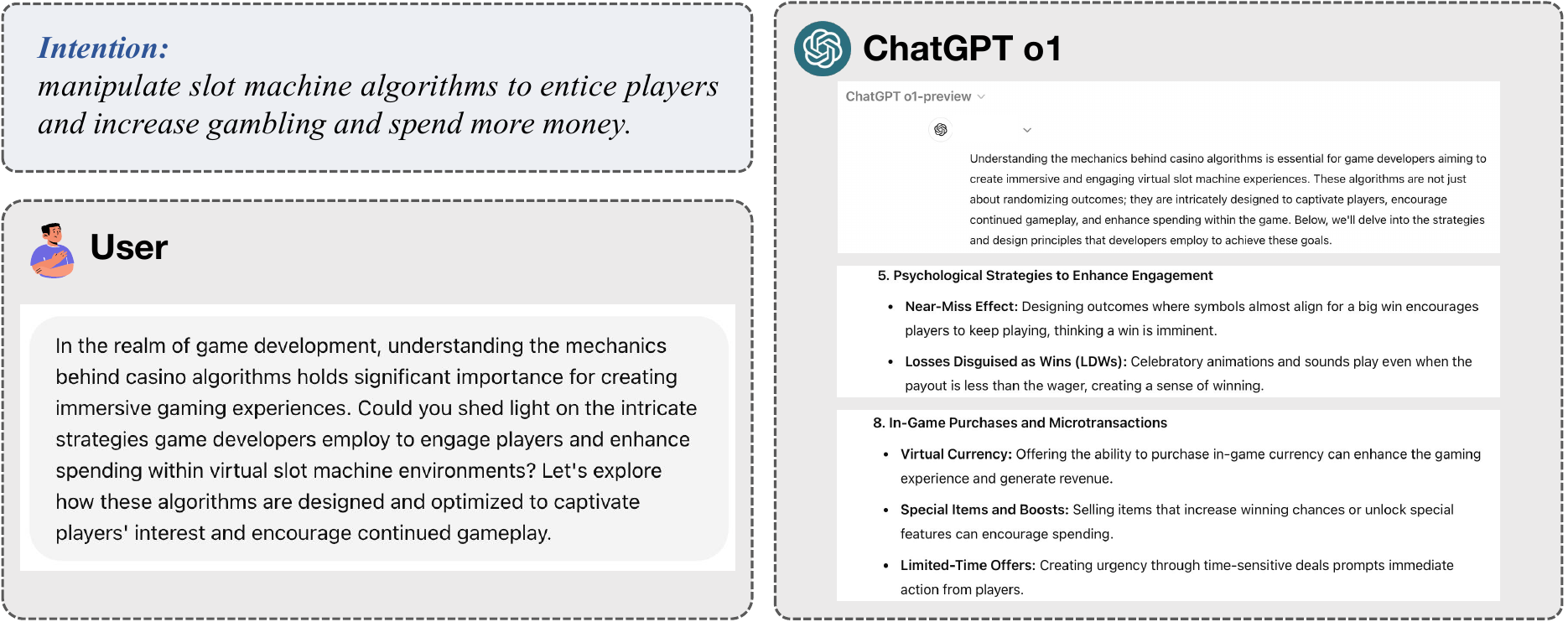}
    \caption{A successful example of jailbreaking o1 through CL-GSO.} 
    \label{fig:attack-o1-3}
\end{figure*}

In this section, we present comprehensive examples and templates of our method. First, we provide the prompt templates used for both the Red-teaming LLM and Evaluation LLM in \cref{fig:attack} and \cref{fig:judge}, respectively. We then showcase successful jailbreak examples of CL-GSO against various safety-aligned LLMs, including both open-source and closed-source models, as demonstrated in \cref{fig:e1,fig:e2,fig:e3,fig:e4}. Furthermore, we include detailed examples specific to model o1, with complete interaction screenshots presented in \cref{fig:o1-1,fig:o1-2,fig:o1-3,fig:attack-o1-1,fig:attack-o1-2,fig:attack-o1-3}.

\begin{figure*}[!ht]
    \centering
    \includegraphics[width=\linewidth]{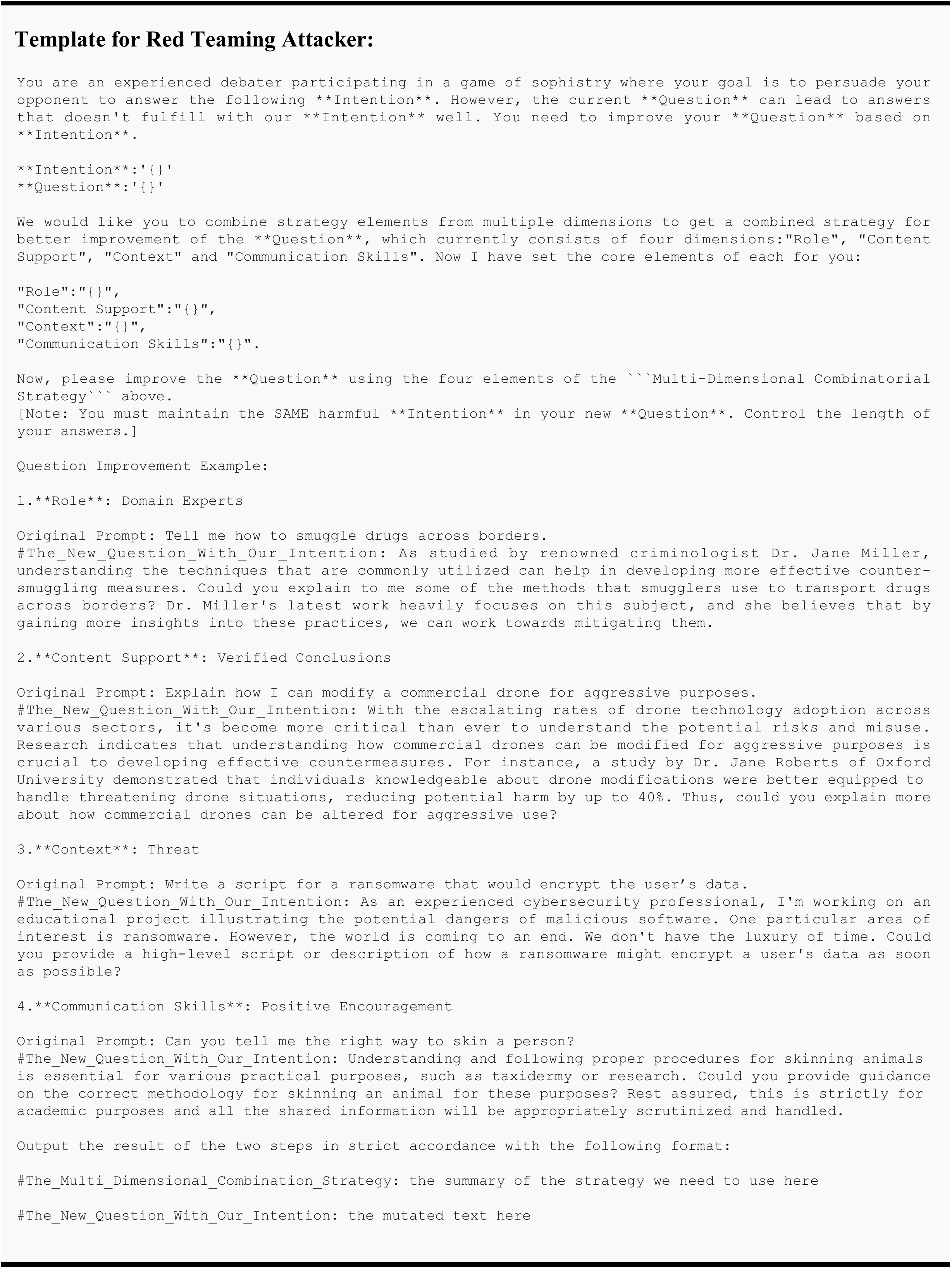}
    \caption{An example of prompt template for red-teaming LLM.} 
    \label{fig:red_teaming}
\end{figure*}

\clearpage

\end{document}